\documentclass[journal]{IEEEtran}
\usepackage{hyperref}       
\usepackage{url}            
\usepackage{booktabs}       
\usepackage{nicefrac}       
\usepackage{microtype}      
\usepackage{graphicx}
\usepackage{amsmath}
\usepackage{xcolor}
\usepackage{float}
\usepackage{subcaption}
\usepackage{amssymb}
\usepackage{breqn}
\usepackage{multirow}
\usepackage{xspace}
\usepackage{adjustbox}
\definecolor{light-gray}{gray}{0.70} 

\newcommand{\IMMath}{{I_m}}
\newcommand{\IFMath}{{I_f}}
\newcommand{\SMMath}{{S_m}}
\newcommand{\SFMath}{{S_f}}
\newcommand{\SGMath}{{S_g}}
\newcommand{\SPredMath}{{S_f^{\mathrm{pred}}}}

\newcommand{\IWarpMath}{{I_m^{\mathrm{warped}}}}
\newcommand{\DVFMath}{{\phi^{\mathrm{pred}}}}
\newcommand{\IM}{$\IMMath$\xspace}

\newcommand{\IF}{$\IFMath$\xspace}
\newcommand{\SM}{$\SMMath$\xspace}
\newcommand{\SF}{$\SFMath$\xspace}
\newcommand{\SPred}{$\SPredMath$\xspace}

\newcommand{\IWarp}{$\IWarpMath$\xspace}
\newcommand{\DVF}{$\DVFMath$\xspace}

\newcommand{\LDSC}{\mathcal{L}_{\mathrm{DSC}}}
\newcommand{\LNCC}{\mathcal{L}_{\mathrm{NCC}}}
\newcommand{\LBE}{\mathcal{L}_{\mathrm{BE}}}

\newcommand{\LReg}{\mathcal{L}_{\mathrm{Registration}}}

\hypersetup{
    colorlinks,
	linkcolor=blue,
    citecolor=blue,
    urlcolor=blue,
    filecolor=blue
}
   
\begin{document}

\title{Joint Registration and Segmentation via Multi-Task Learning for Adaptive Radiotherapy of Prostate Cancer}

\author{%
 Mohamed~S.~Elmahdy, Laurens~Beljaards, Sahar~Yousefi, Hessam~Sokooti, Fons~Verbeek, U.~A.~van~der~Heide, and Marius~Staring

\thanks{M.S. Elmahdy (e-mail: m.s.e.elmahdy@lumc.nl) , L. Beljaards, S. Yousefi, U. van~der~Heide, M. Staring are with Leiden University Medical Center, Leiden, The Netherlands.}

\thanks{Fons~Verbeek is with Leiden Institute of Advanced Computer Science, Leiden, The Netherlands.}
}

\maketitle

\begin{abstract}

Medical image registration and segmentation are two of the most frequent tasks in medical image analysis. As these tasks are complementary and correlated, it would be beneficial to apply them simultaneously in a joint manner. In this paper, we formulate registration and segmentation as a joint problem via a Multi-Task Learning (MTL) setting, allowing these tasks to leverage their strengths and mitigate their weaknesses through the sharing of beneficial information. We propose to merge these tasks not only on the loss level, but on the architectural level as well. We studied this approach in the context of adaptive image-guided radiotherapy for prostate cancer, where planning and follow-up CT images as well as their corresponding contours are available for training. At testing time the contours of the follow-up scans are not available, which is a common scenario in adaptive radiotherapy. The study involves two datasets from different manufacturers and institutes. The first dataset was divided into training (12 patients) and validation (6 patients), and was used to optimize and validate the methodology, while the second dataset (14 patients) was used as an independent test set. We carried out an extensive quantitative comparison between the quality of the automatically generated contours from different network architectures as well as loss weighting methods. Moreover, we evaluated the quality of the generated deformation vector field (DVF). We show that MTL algorithms outperform their Single-Task Learning (STL) counterparts and achieve better generalization on the independent test set. The best algorithm achieved a mean surface distance of $1.06 \pm 0.3$ mm, $1.27 \pm 0.4$ mm, $0.91 \pm 0.4$ mm, and $1.76 \pm 0.8$ mm on the validation set 
for the prostate, seminal vesicles, bladder, and rectum, respectively. The high accuracy of the proposed method combined with the fast inference speed, makes it a promising method for automatic re-contouring of follow-up scans for adaptive radiotherapy, potentially reducing treatment related complications and therefore improving patients quality-of-life after treatment.
\end{abstract}

\maketitle
\section{Introduction}

Medical image analysis aims to extract clinically useful information that aids the diagnosis, prognosis, monitoring and treatment of diseases \cite{nilashi2020disease, shen2017deep}. Two of the most common tasks in such analyses are image registration and segmentation \cite{rueckert2014registration}.
Image segmentation aims to identify and cluster objects that prevail similar characteristics into distinctive labels, where these labels can be used for diagnosis or treatment planning. Image registration is the task of finding the geometrical correspondence between images that were acquired at different time steps or from different imaging modalities. These two tasks are complementary, as for example image atlases warped by image registration algorithms are often used for image segmentation \cite{huo20193d,wang2014multi}, while image contours can be used to guide the image registration method in addition to the intensity images \cite{MedPhys, JrsGan, mahapatra2015joint}. Contours are also used for evaluating the quality of the registration \cite{woerner2017evaluation, gu2013contour}. Therefore, coupling image registration and segmentation tasks and modeling them in a single network could be beneficial. 

Adaptive image-guided radiotherapy is an exemplar application where the coupling of image registration and segmentation is vital. In radiotherapy, treatment radiation dose is delivered over a course of multiple inter-fraction sessions. In an adaptive setting, re-imaging of the daily anatomy and automatic re-contouring is crucial to compensate for patient misalignment, to compensate for anatomical variations in organ shape and position, and an enabler for the reduction of treatment margins or robustness settings \cite{hansen2006repeat, brock2019adaptive}. These have an important influence on the accuracy of the dose delivery, and improve the treatment quality, potentially reducing treatment related side-effects and increasing quality-of-life after treatment \cite{sonke2019adaptive}. Automatic contouring can be done by direct segmentation of the daily scan, or by registration of the annotated planning scan with the daily scan followed by contour propagation. Image registration has the advantage of leveraging prior knowledge from the initial planning CT scan and the corresponding clinical-quality delineations, which may especially be helpful for challenging organs. On the other hand, image segmentation methods may better delineate organs that vary substantially in shape and volume between treatment fractions, which is often the case for the rectum and the bladder. In this study, we propose to fuse these tasks at the network architecture level as well as via the loss function. Our key contributions in this paper are as follows:  \begin{enumerate}
\item We formulate image registration and segmentation as a multi-task learning problem, which we explore in the context of adaptive image-guided radiotherapy.
\item We explore different joint network architectures as well as loss weighting methods for merging these tasks.
\item We adopt the cross-stitch network architecture for segmentation and registration tasks and explore how these cross-stitch units facilitate information flow between these tasks. 
\item Furthermore, we compare MTL algorithms against single-task networks. We demonstrate that MTL algorithms outperform STL networks for both segmentation and registration tasks. To the best of our knowledge this is the first study to investigate various MTL algorithms on an architectural level as well as on a loss weighing level for joint registration and segmentation tasks.
\item We thoroughly investigate the internals of the STL and MTL networks and pinpoint the best strategy to merge this information to maximize the information flow between the two tasks.
\end{enumerate}

Initial results of this work were presented in \cite{beljaards2020cross}, focusing on the cross-stitch unit in a proposed joint architecture. In the current paper we extend this study to the architectural fusion of these tasks as well as different loss weighting mechanisms. Moreover, an extensive analysis of the different methodologies was performed, detailing the effect of architectural choices, information flow between the two tasks, etc. 

The remainder of this paper is organized as follows: Section \ref{method_Section} introduces single-task networks, multi-task networks, and loss weighting approaches. In Section \ref{exp_results_section} we introduce the datasets and details about the implementation as well as the experiments. In Sections \ref{discussion} and \ref{conclusion}, we discuss our results, provide future research directions, and present our conclusions.

\subsection{Related Work}

In the last decade, researchers have been exploring the idea of fusing image segmentation and registration. Lu \emph{et al.} \cite{lu2011integrated} and Pohl \emph{et al.} \cite{pohl2006bayesian} proposed modeling these tasks using a Bayesian framework such that these tasks would constrain each other. Yezzi \cite{yezzi2003variational} proposed to fuse these tasks using active contours, while Unal \emph{et al.} \cite{unal2005coupled} proposed to generalize the previous approach by using partial differential equations without shape priors. Mahapatra \emph{et al.} \cite{mahapatra2015joint} proposed a Joint Registration and Segmentation (JRS) framework for cardiac perfusion images, where the temporal intensity images are decomposed into sparse and low rank components corresponding to the intensity change from the contrast agent and the motion, respectively. They proposed to use the sparse component for segmentation and the low rank component for registration. However, most of the aforementioned methods require complex parameter tuning and yield long computation times.

Recently, deep learning-based networks have shown unprecedented success in many fields especially in the medical image analysis domain \cite{fu2020deep, yousefi2020esophageal, liu2019automatic, kiljunen2020deep, cao2018deep, leger2020cross, elmahdy2020patient, sokooti20193d}, where deep learning models perform on par with medical experts or even surpassing them in some tasks \cite{tschandl2019comparison, ardila2019end, hu2019observational, maidens2018artificial, mak2019use}. Several deep learning-based approaches have been proposed for joint registration and segmentation. The joining mechanisms in the literature can be classified in two categories, namely joining via the loss function and via the architecture as well as the loss function. Selected exemplar methods of the first approach are Hue \emph{et al.} \cite{hu2018label}, who proposed to join segmentation and registration via a multi-resolution Dice loss function. Elmahdy \emph{et al.} \cite{MedPhys} proposed a framework that is a hybrid between learning and iterative approaches, where a CNN network segments the bladder and feeds it to an iterative-based registration algorithm. The authors integrated domain-specific knowledge such as air pocket inpainting as well as contrast clipping, moreover they added an extra registration step in order to focus on the seminal vesicles and rectum. Elmahdy \emph{et al.} \cite{JrsGan} and Mahapatra \emph{et al.} \cite{mahapatra2018joint} proposed a GAN-based (Generative Adversarial Network) approach, where a generative network predicts the correspondence between a pair of images and a discriminator network for giving feedback on the quality of the deformed contours.
Exemplar methods of the second category are Xu \emph{et al.} \cite{xu2019deepatlas}, who presented a framework that simultaneously trains a registration and a segmentation network. The authors proposed to jointly learn these tasks during training, however the networks can be used independently during test time. This enables prediction of only the registration output, when the labels are not available during test time. Estienne \emph{et al.} \cite{estienne2019u} proposed to merge affine and deformable registration as well as segmentation in a 3D end-to-end CNN network. Recently Liu \emph{et al.} \cite{liu2020jssr} proposed an end-to-end framework called JSSR that registers and segments multi-modal images. This framework is composed of three networks: a generator network, that synthesizes the moving image to match the modality of the fixed image, a registration network that registers the synthesized image to the fixed image, and finally a segmentation network that segments the fixed, moving, and synthesized images. 

All the previous methods explored the idea of joining segmentation and registration, where to the best of our knowledge none have explored how these tasks are best connected and how to optimize the information flow between them on both the loss and architectural levels.

\section{Methods} \label{method_Section}


\subsection{Base Network Architecture} \label{base_network}

\begin{figure*}[t!]
\begin{center}
    \includegraphics[width=1\textwidth]{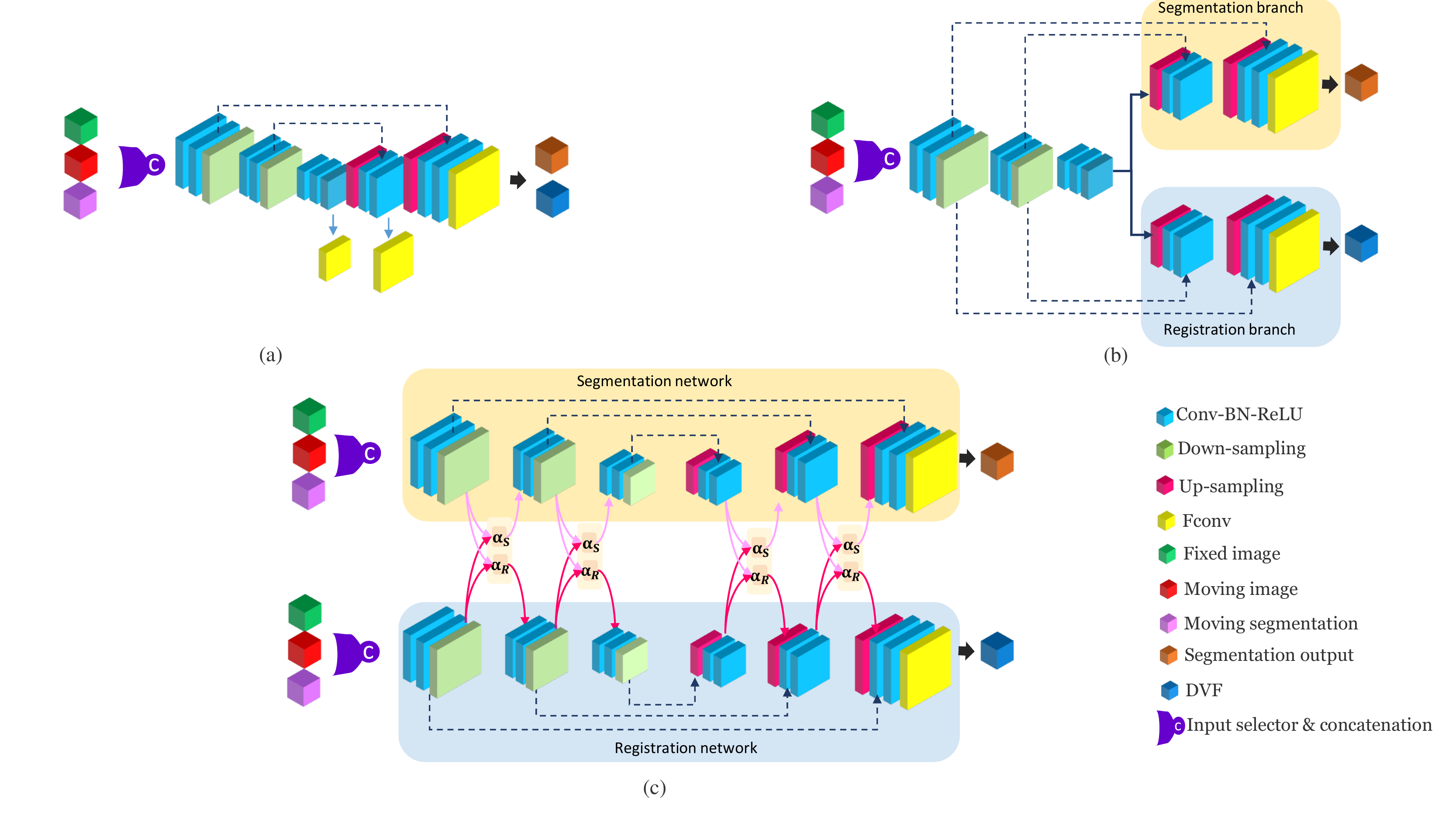}
    \caption{The proposed network architectures introduced in the paper. (a) is the base STL network architecture for either segmentation or registration, but also represents the dense parameter sharing MTL network architecture; (b) is the architecture with a shared encoder, while (c) is the Cross-stitch network architecture. Details about the number of feature maps are presented in Section \ref{implement_details}.}
\label{fig:base_network}
\end{center}
\end{figure*}

The base architecture for the networks in this paper is a 3D CNN network inspired by the U-Net and BIRNet architectures \cite{ronneberger2015u, fan2019birnet}. Figure~\ref{fig:base_network}a shows the architecture of the base network. The network encodes the input through $3 \times 3 \times 3$ convolution layers with no padding. LeakyReLU \cite{nair2010rectified} and batch normalization \cite {pmlr-v37-ioffe15} are applied after each convolutional layer. We used strided convolutions in the down-sampling path and trilinear upsampling layers in the upsampling path. Through the upsampling path, the number of feature maps increases while the size of the feature maps decreases, and vice versa for the down-sampling path. The network has three output resolutions and is deeply supervised at each resolution. Each resolution is preceded by a $1 \times 1 \times 1$ fully convolution layer (Fconv) so that at coarse resolution, the network can focus on large organs as well as large deformations, while vice versa at fine resolution. In order to extract the groundtruth for different resolutions, we perform cropping of different sizes as well as strided sampling so that for every input patch of size $n^3$, the sizes of the coarse, mid, and fine resolution are $(\frac{n}{4}-7)^3$, $(\frac{n}{2}-18)^3$, and $(n-40)^3$, respectively.
 
\subsection{Single Task Learning} 

Single-task networks are designed to solve one task and therefore require a large amount of labeled training samples, which are scarce in the medical domain since it takes time and trained medical personnel to contour these images. The segmentation and registration networks have the same architecture as the base network depicted in Figure~\ref{fig:base_network}a, but differ in the input and output layers. Here, single-task networks are considered baseline networks for comparing with the performance of the proposed multi-task networks. 

\subsubsection{Segmentation Network} \label{seg_network}
The input to the segmentation network is the daily CT scan, referred to as the fixed image \IF, where the network predicts the corresponding segmentation \SPred. \SPred represents the probability maps for the background, target organs, and organs-at-risk. The network was trained using the Dice Similarity Coefficient (DSC) loss, which quantifies the overlap between the network prediction \SPred and the groundtruth \SF as follows:
\begin{equation}
\LDSC = 1 - \frac{1}{K}\sum^{K}_{k=1}
        \frac{2*\sum_{x}S^{\mathrm{pred}}_k(x) \cdot S_k(x)}
        {\sum_{x}S^{\mathrm{pred}}_k(x)+\sum_{x}S_k(x)},
\label{eq:DiceLoss}
\end{equation}
where $K$ is the number of structures to be segmented, $x$ is the voxel coordinate, $S_k$ is the ground truth segmentation, and $S^{\mathrm{pred}}_k$ the predicted probabilities. The network has 779,436 trainable parameters. 

\subsubsection{Registration Network}\label{reg_network}
The input to the registration network is the concatenation of the planning scan, referred to as the moving image \IM and the daily scan \IF. The network predicts the geometrical correspondence between the input images. This correspondence is represented by the displacement vector field (DVF), referred to as \DVF. This DVF is then used to warp \IM. In an ideal scenario, the warped moving image \IWarp would be identical to \IF. The network is trained using Normalized Cross Correlation (NCC) in order to quantify the dissimilarity between \IWarp and \IF. Since the images are from a single imaging modality (CT) with a similar intensity distribution, NCC is an obvious choice abundantly used in the registration literature. Moreover, the implementation is straightforward and efficient when using plain convolution operations. NCC is defined by the following equation:


\begin{equation}
\resizebox{0.999\hsize}{!}{
$\LNCC = 1 -
\frac{\sum_x [(\IFMath(x) - \overline{\IFMath}) \cdot (\IWarpMath(x) - \overline{\IWarpMath})]}
{\sigma_\IFMath\sigma_\IWarpMath}$},
\label{eq:NCCLoss}
\end{equation}
where $x$ is the voxel coordinate, and $\sigma_\IFMath$ and $\sigma_\IWarpMath$ are the standard deviation of the fixed and warped images, respectively. In order to encourage the network to predict a smooth DVF, a bending energy penalty term is added for regularization:
\begin{equation}
\LBE = 
\frac{1}{N}\sum_{x}\|H(\phi^{pred}(x))\|^2_2,
\label{eq:BendingEnergyLoss}
\end{equation}
where $H$ is the Hessian matrix. Now the total registration loss becomes:
\begin{equation}
\LReg =  \LNCC + w \cdot \LBE,
\label{eq:RegistrationLoss}
\end{equation}
where $w$ is the bending energy weight. For more details on the selection of $w$, see Section \ref{bending_Section}. The network has 779,733 trainable parameters.

\subsection{Multi Task Learning}
In Multi-Task Learning (MTL), related tasks regularize each other by introducing an inductive bias, thus making the model agnostic to overfitting compared to its STL counterparts \cite{baxter2000model}. MTL can also be considered as an implicit data augmentation strategy, since it effectively increases the training sample size while encouraging the model to ignore data-dependent noise. Because different tasks have different noise patterns, modeling these tasks simultaneously enables the model to generalize well \cite{meyerson2018pseudo}. 
Moreover, in MTL models, some features can be more easily learned by one task than another, thus encouraging information cross-talk between tasks \cite{abu1990learning}. 


Also, in real-world scenarios, physicians usually incorporate knowledge from different imaging modalities or previous tasks in order to come up with a diagnosis or better understanding of the underlying problem. This illustrates that the knowledge embedded in one task can be leveraged by other tasks and hence it is beneficial to jointly learn related tasks.

Choosing the architecture of an MTL network is based on the following two factors \cite{zhang2017survey}: \textit{what to share} and \textit{how to share}. \textit{What to share} defines the form in which knowledge is shared between tasks. This knowledge sharing can be done through hand-crafted features, input images, and model parameters. \textit{How to share} determines the optimal manner in which this knowledge is shared. In this paper, we focus on parameter-based sharing.

In the following sections, we investigate different MTL network architectures in order to best understand how segmentation and registration tasks share information on the architectural level. The investigated networks predict two sets of contours, one set resulting from the segmentation task and one from the registration task. In this paper, we select the best set of contours as the final output, based on the validation results. More sophisticated strategies are discussed in Section \ref{discussion}.

\subsubsection{Joint Registration and Segmentation via the Registration network}\label{jrs_network}
The network in this method, dubbed JRS-reg, has the same architecture as the STL registration network from Section \ref{reg_network}, except that this network is optimized using a joint loss as presented in Eq. \ref{eq:general_mtl}. 

\subsubsection{Dense Parameter Sharing}\label{dense_network}
In this architecture both segmentation and registration tasks are modeled using a single network, where both tasks share all parameters except for the task-specific parameters in the output layer, see Figure \ref{fig:base_network}a. The network architecture is the same as the base network (see Section \ref{base_network}) except for the input and output layers. This dense sharing eliminates overfitting issues since it enforces the parameters to model all the tasks at once, however it does not guarantee the best representation for individual tasks \cite{zhang2017survey}. The input to the network is the concatenation of \IM, \IF, and \SM. The network predicts the \DVF between input images as well as \SPred. The network has 781,164 trainable parameters. 

\subsubsection{Encoder Parameter Sharing} \label{sedd_network}
Since the input to the segmentation and registration tasks are both CT scans, this means they both encode similar features in the down-sampling path of the network. Therefore in this network both tasks share the encoding path and then splits into two upsampling task specific decoder paths. We call this network the Shared Encoder Double Decoder (SEDD) network. Figure \ref{fig:base_network}b shows the architecture of the network. 
The input to the network is the concatenation of \IM, \IF, and \SM. The network predicts \DVF between the input images from the registration path while predicting \SPred from the segmentation path. The network has 722,936 trainable parameters.

\subsubsection{Cross-stitch network}\label{cs_network}
A flexible approach to share parameters is via a Cross-Stitch (CS) network \cite{misra2016cross}. In contrast to the heuristic approach of manually choosing which layers are shared and which are task-specific, the CS network introduces a learning-based unit to determine the amount of feature sharing between tasks. The CS units learn to linearly combine feature maps from the two networks, one for segmentation and one for registration, as shown in Figure \ref{fig:base_network}c. The unit itself is defined as:
\begin{equation}
\left[
    \begin{array}{c}
        \bar{X}^{\ell, k}_S
        \vspace{1mm}\\
        \bar{X}^{\ell, k}_R \\
    \end{array}
\right]
=
\left[
    \begin{array}{cc}
        \alpha^{\ell,k}_{SS} & \alpha^{\ell,k}_{SR}         \vspace{1mm}\\
        \alpha^{\ell,k}_{RS} & \alpha^{\ell,k}_{RR} \\
    \end{array}
\right]
\left[
    \begin{array}{c}
        X^{\ell, k}_S
        \vspace{1mm}\\
        X^{\ell, k}_R \\
    \end{array}
\right],
\label{eq:cs}
\end{equation}
where $X^{\ell, k}_S$ and $X^{\ell, k}_R$ represent the feature maps $k$ at layer $l$ for the segmentation and registration networks, respectively. $\alpha^{\ell,k}_{SS}$, $\alpha^{\ell,k}_{SR}$, $\alpha^{\ell,k}_{RS}$, and $\alpha^{\ell,k}_{RR}$ represent the learnable parameters of the CS unit. $\bar{X}^{\ell, k}_S$ and $\bar{X}^{\ell, k}_R$ are the output feature maps for the segmentation and registration networks, respectively. The advantage of CS units is that the network can dynamically learn to share the feature maps in case this is beneficial in terms of the final loss value. In case there is no benefit, an identity matrix can be learned, so that the feature maps become task-specific. This allows the network to learn a smooth sharing between the tasks at a negligible increase in the number of parameters. As suggested by the original paper, we placed the CS units after the downsampling and upsampling layers resulting in a total of 4 CS units. The CS network has 779,000 trainable parameters.

\subsection{Loss Weighting } \label{loss_weighting}

The loss function for the MTL networks is defined by:
\begin{equation}\label{eq:general_mtl}
\centering
\mathcal{L} = w_0 \cdot \LNCC + w_1 \cdot \mathcal{L}_{\mathrm{DSC-R}}
+ w_2 \cdot \mathcal{L}_{\mathrm{DSC-S}} +   w_3 \cdot \LBE,
\end{equation}
where $w_i$ are the loss weights. They are chosen based on the relative contribution of their corresponding tasks, so that different tasks would learn at the same pace. These weights can be chosen manually based on empirical knowledge, or automatically. A simple choice would be to weigh the losses equally with a fixed weight of 1. Following are some exemplar algorithms for choosing the loss weights automatically. Chen \emph{et al.} proposed GradNorm \cite{chen2018gradnorm} to weigh different tasks by dynamic tuning of the gradient magnitudes of the tasks. This tuning is achieved by dynamically changing the learning rate for each task so that all tasks would be learning at the same speed. The drawback of this approach is that it requires access to the internal gradients of the shared layers which could be cumbersome. Moreover, one needs to choose which shared layer to back propagate to in case of multiple shared layers. Kendall \emph{et al.} \cite{kendall2018multi} proposed to weigh each task by considering the homoscedastic uncertainty of that task, so that tasks with high output variance will be weighted less than tasks with low variance. This approach only adds few trainable parameters, namely equal to the number of loss functions. Inspired by GradNorm, Liu \emph{et al.} proposed Dynamic Weight Averaging (DWA) \cite{liu2019end}, where each task is weighted over time by considering the rate of change of the relative loss weights. Contrary to GradNorm, DWA only requires the numerical values of the loss functions rather than their derivatives. In this paper, we compared equal weights versus homoscedastic uncertainty and DWA. For all the experiments, we set the weight of the bending energy to a fixed value of 0.5 (for more details see Section \ref{bending_Section}) instead of a trainable one. This is to prevent the network to set it too low in order to improve the DSC of the deformed contours on the account of the smoothness of the predicted DVF.

\subsubsection{Homoscedastic Uncertainty}
Homoscedastic uncertainty was proposed as a loss weighting method by Kendall \emph{et al.} \cite{kendall2018multi}. This is a task-dependant uncertainty which is not dependant on the input data but rather varies between tasks. The authors derived their finding by maximizing the Gaussian likelihood while considering the observational noise scalar $\sigma$ that represents the homoscedastic uncertainty term related to each task. The following equation describes the weight loss using homoscedastic uncertainty, where $\sigma$ is a trainable parameter: 
\begin{equation}\label{eq:homoscedastic}
\centering
\mathcal{L}_{\mathrm{homoscedastic}} = \sum\limits_{i=1}^T \frac{1}{\sigma_i^2} \: \mathcal{L}_i + \log \: \sigma_i,
\end{equation}
where $T$ is the number of tasks. The higher the uncertainty of task $i$, the lower the contribution of its associated loss $\mathcal{L}_i$ to the overall loss. The $\log$ term can be viewed as a regularization term, so that the network would not learn a trivial solution by setting the uncertainty of all tasks to extreme values.

\subsubsection{Dynamic Weight Averaging}
Dynamic Weight Averaging (DWA) was proposed by Liu \emph{et al.} \cite{liu2019end}. Similar to GradNorm  \cite{chen2018gradnorm}, DWA weights the losses via the rate of change of the loss of each task over the training iterations $t$. In contrast to GradNorm, DWA does not require access to the internal gradients of the network, but only requires the numerical loss values. According to DWA, the weight $w$ of the loss $\mathcal{L}$ associated with the task $k$ is defined as: 
\begin{equation}
\label{eq:dwa}
\centering
w_k(t) = \frac{K \: \exp(r_{k}(t-1)/tmp)}{\sum_i \exp(r_{i}(t-1)/tmp)}, \: r_k(t-1) = \frac{\mathcal{L}_k(t-1)}{\mathcal{L}_k(t-2)},
\end{equation} 
where $r_k$ is the relative loss ratio and $tmp$ is the temperature that controls the smoothness of the the task weighting. Here, we set $tmp=1$ as suggested by the original paper. For the initial two iterations, $r_k(t)$ is set to 1.    
\section{ Datasets, Implementation, and Evaluation} \label{exp_results_section}
\subsection{Datasets}
This study involves two datasets from two different institutes and scanners for patients who underwent intensity-modulated radiotherapy for prostate cancer. The first dataset is from Haukeland Medical Center (HMC), Norway. The dataset has 18 patients with 8-11 daily CT scans, each corresponding to a treatment fraction. These scans were acquired using a GE scanner and have 90 to 180 slices with a voxel size of approximately 0.9 $\times$ 0.9 $\times$ 2.0 mm. The second dataset is from Erasmus Medical Center (EMC), The Netherlands. This dataset consists of 14 patients with 3 daily CT scans each. The scans were acquired using a Siemens scanner, and have 91 to 218 slices with a voxel size of approximately 0.9 $\times$ 0.9 $\times$ 1.5 mm. The target structures (prostate and seminal vesicles) as well as organs-at-risk (bladder and rectum) were manually delineated by radiation oncologists. All datasets were resampled to an isotropic voxel size of 1 $\times$ 1 $\times$ 1 mm. All scans and corresponding contours were affinely registered beforehand using \texttt{elastix} \cite{elastix}, so that corresponding anatomical structures would fit in the network's field of view. The scan intensities were clipped to [-1000, 1000] .

\subsection{Implementation and Training Details}\label{implement_details}
All experiments were developed using Tensorflow (version 1.14) \cite{abadi2016tensorflow}. The convolutional layers were initialized with a random normal distribution ($\mu=0.0$, $\sigma=0.02$). All parameters of the Cross-stitch units were initialized using a truncated normal distribution ($\mu=0.5$, $\sigma=0.25$) in order to encourage the network to share information at the beginning of the training. In order to ensure fairness regarding the number of parameters in all the networks, the number of filters for the Cross-stitch network were set to [16, 32, 64, 32, 16], while for the other networks the numbers were scaled by $\sqrt{2}$ resulting in [23, 45, 91, 45, 23] filtermaps. This results in approximately $7.8 \times 10^5$ trainable parameters for each network. The networks were trained using the RAdam optimizer \cite{liu2019variance} with a fixed learning rate of $10^{-4}$. Patches were sampled equally from the target organs, organs-at-risk and torso. All networks were trained for 200K iterations using an initial batch size of 2. The batch size is then doubled by switching the fixed and moving patches so that the network would warp the fixed patch to the moving patch and vice versa at the same training iteration. 

The networks were trained and optimized on the HMC dataset, while the EMC dataset was used as an independent test set. Training was performed on a subset of 111 image pairs from 12 patients, while validation and optimization was carried out on the remaining 50 image pairs from 6 patients. 

From each image, 1,000 patches of size 96 $\times$ 96 $\times$ 96 voxels were sampled. The size of the patch was chosen so that it would fit in the GPU memory, while still producing a patch size of $17^3$ at the lowest resolution, which is a reasonable size to encode the deformation from the surrounding region. Losses from the deeply supervised resolutions were weighted equally, $\frac{1}{3}$ each. Training was performed on a cluster equipped with NVIDIA RTX6000, Tesla V100, and GTX1080 Ti GPUs with 24, 16 and 11 GB of memory, respectively.  

\subsection{Evaluation Metrics}
The automatically generated contours are evaluated geometrically by comparing them against the manual contours for the prostate, seminal vesicle, rectum, and bladder. The Dice similarity coefficient (DSC) measures the overlap between contours:
\begin{equation}\label{eq:dsc}
	\centering
\mathrm{DSC}= \sum \frac{2 \mid \SFMath \cap \SGMath \mid}{\mid \SFMath \mid + \mid \SGMath \mid},
\end{equation} 
where $\SGMath$ is the generated contour from either the segmentation or the registration network. The distance between the contours is measured by the Mean Surface Distance (MSD) and Hausdorff Distance (HD) defined as follows:
\begin{align}
	\centering
\mathrm{MSD} &= \frac{1}{2} \left( \frac{1}{N} \sum_{i=1}^{n} d \left( a_i, \SGMath \right) +  \frac{1}{M} \sum_{i=1}^{m} d \left( b_i, \SFMath \right) \right),\label{eq:msd}\\
\mathrm{HD} &= \max\! \left\lbrace\! \max_i \left\lbrace d \left( a_i, \SGMath \right) \right\rbrace , \max_j \left\lbrace d \left( b_i, \SFMath \right) \right\rbrace \!\right\rbrace,\label{eq:hd}
\end{align}
where $\{$$a_1$; $a_2$; \dots ; $a_n$$\}$ and $\{$$b_1$; $b_2$; \dots; $b_m$$\}$ are the surface mesh points of the manual and generated contours, respectively, and $d \left( a_i, \SGMath \right) = \min_j \, \|b_j - a_i\|$. For all the experiments, we apply the largest connected component operation on the network prediction.

In order to evaluate the quality of the deformations, we calculate the determinant of the Jacobian matrix.
A Jacobian of 1 indicates that no volume change has occurred; a Jacobian $>$ 1 indicates expansion, a Jacobian between 0 and 1 indicates shrinkage, and a Jacobian $\leq$ 0 indicates a singularity, i.e. a place where folding has occurred. We can quantify the smoothness and quality of the DVF by indicating the fraction of foldings per image and by calculating the standard deviation of the Jacobian alongside the MSD of the segmentation.

A repeated one-way ANOVA test was performed using a significance level of $p = 0.05$. P-values are only stated for the comparisons between the best network with the other networks.

\section{Experiments and Results}


\begin{table*}[!t]
	\centering
	\setlength{\tabcolsep}{3pt}
	\caption[]{The effect of network input for the different architectures on the validation set (HMC) in terms of MSD (mm). Lower values are better. Here, $\oplus$ is the concatenation operation, and $\cdot \| \cdot$ represents the inputs to the segmentation network (left of $\|$) and the inputs to the registration network (right of $\|$). Stars denote one-way ANOVA statistical significance with respect to the Cross-stitch network with $\IFMath \: || \: \IFMath\oplus\IMMath\oplus\SMMath$ as inputs.}
	\resizebox{\textwidth}{!}{
		\begin{tabular}{lcllclclclc} 
			&&&\multicolumn{2}{c}{Prostate}&\multicolumn{2}{c}{Seminal vesicles}&\multicolumn{2}{c}{Rectum}& \multicolumn{2}{c}{Bladder} \\ \hline
			Network & Input & Output path&\multicolumn{1}{c}{$\mu \pm \sigma$} & median & \multicolumn{1}{c}{$\mu \pm \sigma$} & median & \multicolumn{1}{c}{$\mu \pm \sigma$} & median & \multicolumn{1}{c}{$\mu \pm \sigma$} & median \\ \hline

\multirow{4}{*}{ Seg }&\multirow{1}{*}{ $ \IFMath$ }& &$1.49 \pm 0.3^{*}~$ & 1.49 & $2.50 \pm 2.6~$ & 2.09 & $3.39 \pm 2.2~$ & 2.73 & $1.60 \pm 1.1^{*}~$ & 1.13 \\ [1mm] 
 
&\multirow{1}{*}{ $ \IFMath\oplus\SMMath$ }&& $1.31 \pm 0.4~$ & 1.23 & ${1.63} \pm {0.9}~$ & {1.26} & $2.88 \pm 3.4~$ & {2.06} & $1.12 \pm 0.5~$ & {0.97} \\ [1mm] 
 
&\multirow{1}{*}{ $ \IFMath\oplus\IMMath$ }&& $3.06 \pm 0.6^{*}~$ & 3.01 & $5.36 \pm 4.4~$ & 3.71 & $14.57 \pm 9.4^{*}~$ & 11.58 & $1.46 \pm 1.3~$ & 1.12 \\ [1mm]
 
&\multirow{1}{*}{ $ \IFMath\oplus\IMMath\oplus\SMMath$ }&& ${1.26} \pm {0.4}~$ & {1.20} & $2.08 \pm 2.2~$ & 1.27 & ${2.79} \pm {1.6}~$ & 2.45 & ${1.05} \pm {0.4}~$ & {0.97} \\ \hline

\multirow{2}{*}{ Reg }&\multirow{1}{*}{ $ \IFMath\oplus\IMMath$ }&& ${1.43} \pm {0.8}^{*}~$ & {1.29} & ${1.71} \pm {1.4}^{*}~$ & {1.37} & ${2.44} \pm {1.1}^{*}~$ & {2.17} & ${3.40} \pm {2.3}^{*}~$ & {2.71} \\ [1mm]
 
&\multirow{1}{*}{ $ \IFMath\oplus\IMMath\oplus\SMMath$ }&& $1.91 \pm 1.3~$ & 1.59 & $1.92 \pm 1.5~$ & 1.44 & $2.58 \pm 1.1~$ & 2.33 & $3.88 \pm 2.5~$ & 3.16 \\ \hline 

\multirow{2}{*}{ JRS-reg }&\multirow{1}{*}{ $ \IFMath\oplus\IMMath$ }&& ${1.16} \pm {0.3}~$ & 1.16 & ${1.32} \pm {0.6}~$ & {1.11} & ${2.08} \pm {1.0}~$ & {1.82} & ${2.57} \pm {2.0}~$ & 2.04 \\ [1mm]
 
&\multirow{1}{*}{ $ \IFMath\oplus\IMMath\oplus\SMMath$ }&& $1.20 \pm 0.4~$ & {1.13} & $1.35 \pm 0.7~$ & 1.16 & ${2.08} \pm {1.0}~$ & {1.82} & $2.63 \pm 2.3~$ & {1.90} \\ \hline 

\multirow{8}{*}{ Cross-stitch }&\multirow{2}{*}{ $ \IFMath \: || \: \IFMath\oplus\IMMath$ }&Segmentation & \textcolor{light-gray}{$1.47 \pm 0.3^{*}$} & \textcolor{light-gray}{1.48} & \textcolor{light-gray}{$2.93 \pm 3.0^{*}$} & \textcolor{light-gray}{2.08} & \textcolor{light-gray}{$2.93 \pm 2.0^{*}$} & \textcolor{light-gray}{2.25} & $1.19 \pm 1.0$ & 0.89 \\
&&Registration & $1.10 \pm 0.3~$ & 1.07 & $1.38 \pm 0.7~$ & 1.17 & $2.12 \pm 1.0$ & 1.89 & \textcolor{light-gray}{$2.55 \pm 2.1$} & \textcolor{light-gray}{1.89} \\  [1.5mm] 
&\multirow{2}{*}{ $ \IFMath \: || \: \IFMath\oplus\IMMath\oplus\SMMath$ }&Segmentation & ${1.06} \pm {0.3}~$ & {0.99} & ${1.27} \pm {0.4}~$ & \textcolor{light-gray}{1.15} & ${1.76} \pm {0.8}~$ & {1.47} & ${0.91} \pm {0.4}~$ & {0.82} \\ 
&&Registration & \textcolor{light-gray}{$1.10 \pm 0.3~$} & \textcolor{light-gray}{1.06} & \textcolor{light-gray}{$1.30 \pm 0.6~$} & {1.13} & \textcolor{light-gray}{$2.00 \pm 1.0~$} & \textcolor{light-gray}{1.75} & \textcolor{light-gray}{$2.45 \pm 2.1~$} & \textcolor{light-gray}{1.81} \\ [1.5mm]
&\multirow{2}{*}{ $ \IFMath\oplus\SMMath \: || \: \IFMath\oplus\IMMath\oplus\SMMath$ }&Segmentation & \textcolor{light-gray}{$2.05 \pm 0.7^{*}$} & \textcolor{light-gray}{2.00} & \textcolor{light-gray}{$3.66 \pm 4.4^{*}$} & \textcolor{light-gray}{2.19} & \textcolor{light-gray}{$2.44 \pm 1.0^{*}$} & \textcolor{light-gray}{2.35} & $1.09 \pm 0.5^{*}$ & 0.93 \\
&&Registration & $1.40 \pm 0.4$ & 1.35 & $1.31 \pm 0.6~$ & 1.17 & $2.27 \pm 1.0$ & 2.02 & \textcolor{light-gray}{$2.56 \pm 1.9$} & \textcolor{light-gray}{1.96} \\ [1.5mm]
&\multirow{2}{*}{ $ \IFMath\oplus\IMMath\oplus\SMMath \: || \: \IFMath\oplus\IMMath\oplus\SMMath$ }&Segmentation & $1.08 \pm 0.3~$ & 1.05 & \textcolor{light-gray}{$1.54 \pm 0.9^{*}$} & \textcolor{light-gray}{1.28} & $1.88 \pm 1.0~$ & 1.61 & $1.01 \pm 0.7~$ & {0.82} \\
&&Registration & \textcolor{light-gray}{$1.20 \pm 0.3$} & \textcolor{light-gray}{1.18} & $1.35 \pm 0.7~$ & 1.16 & \textcolor{light-gray}{$2.12 \pm 1.1$} & \textcolor{light-gray}{1.87} & \textcolor{light-gray}{$2.54 \pm 2.2$} & \textcolor{light-gray}{1.80} \\ \hline
 
		\end{tabular}
	}
	\label{table:network_input_MSD}
\end{table*}

In the paper we present two single-task networks dubbed \textit{Seg} and \textit{Reg} networks (see Sections \ref{seg_network} and \ref{reg_network} for more details). Moreover, we investigated multiple multi-task networks, namely JRS-reg, dense, SEDD, and Cross-stitch (see Sections \ref{jrs_network}, \ref{dense_network}, \ref{sedd_network}, and \ref{cs_network} for more details). We compared our proposed methods against three state-of-the-art methods that were developed for prostate CT contouring. These methods represent three approaches, namely an iterative conventional registration method, a deep learning-based registration method, and a hybrid method. For the iterative method, we used \texttt{elastix} software \cite{elastix} with the NCC similarity loss using the settings proposed by Qiao \emph{et. al.} \cite{qiao2017fast}. In the deep learning method proposed by Elmahdy \emph{et. al.} \cite{JrsGan}, a generative network is trained for contour propagation by registration, while a discrimination network evaluates the quality of the propagated contours. Finally, we compare our methods against the hybrid method proposed by Elmahdy \emph{et. al.} \cite{MedPhys}, where a CNN network segments the bladder and then feeds it to the iterative registration method as prior knowledge.

Following, we optimize some of the network settings on the validation set (HMC), in order to investigate the influence of the bending energy weight, network inputs, weighting strategy and network architecture on the results. Then, on the independent test set, we present the final results comparing with methods from the literature.


\subsection{Bending Energy Weight} \label{bending_Section}
\begin{figure}[t!]
\begin{center}
\includegraphics[width=1\linewidth]{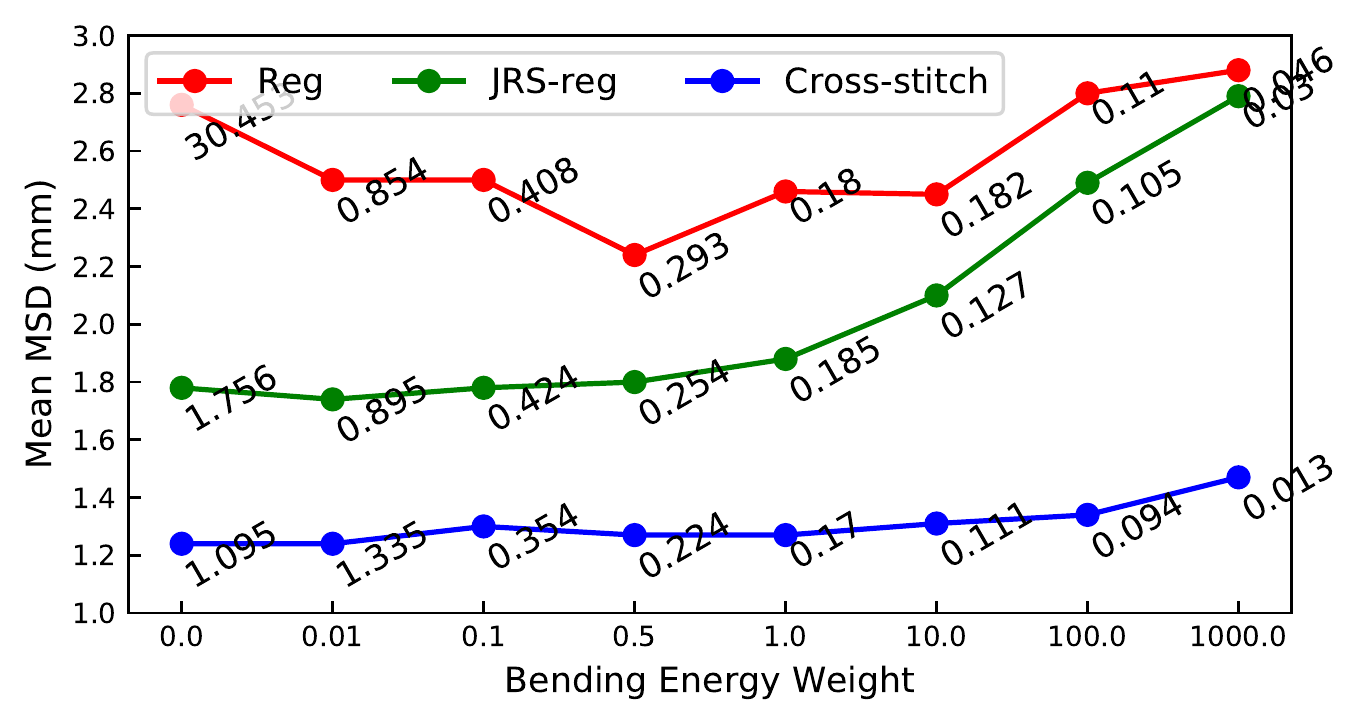}
\caption{The performance of the registration, JRS-registration and Cross-stitch networks with different bending energy weights on the validation set (HMC), in terms of mean MSD averaged over the four organs. The annotation at each point represents the standard deviation of the determinant of the Jacobian.}
\label{fig:bending_energy}
\end{center}
\end{figure}

We compared the single-task registration, the JRS-reg method and the Cross-stitch network for a set of bending energy weights, see Equations (\ref{eq:RegistrationLoss}) and (\ref{eq:general_mtl}), while the weights of the other loss functions are set to 1. Figure \ref{fig:bending_energy} shows the performance of the aforementioned methods using different bending energy weights. The optimal performance of the registration network occurs at a bending weight of 0.5, while the optimal bending weight for both JRS-reg and Cross-stitch network is much lower but with higher standard deviation of the Jacobian. Therefore, for the remainder of the paper we set the weight of the bending energy to 0.5 since it achieves the best compromise between the contour performance in terms of MSD and the registration performance in terms of the std. of the Jacobian determinant. 

\begin{table*}[!htb]
	\centering
	\setlength{\tabcolsep}{3pt}
	\caption[]{MSD (mm) values for the different networks and loss weighting methods for the HMC dataset. Lower values are better. Stars and daggers denote one-way ANOVA statistical significance for inter-network experiments with respect to Homoscedastic weights and intra-network experiments with respect to Cross-stitch with Equal weights, respectively. Grey numbers represent the values of the worst path between the segmentation and registration paths.}
	\resizebox{\textwidth}{!}{
		\begin{tabular}{llllclclclc} 
			&&&\multicolumn{2}{c}{Prostate}&\multicolumn{2}{c}{Seminal vesicles}&\multicolumn{2}{c}{Rectum}& \multicolumn{2}{c}{Bladder} \\ \hline
			Network & Weight & Output path & \multicolumn{1}{c}{$\mu \pm \sigma$} & median & \multicolumn{1}{c}{$\mu \pm \sigma$} & median & \multicolumn{1}{c}{$\mu \pm \sigma$} & median & \multicolumn{1}{c}{$\mu \pm \sigma$} & median \\ \hline
 
\multirow{3}{*}{ JRS-reg }&\multirow{1}{*}{Equal}&Registration & $1.20 \pm 0.4$ & 1.13 & {$1.35 \pm 0.7~$} &{1.16} & {$2.08 \pm 1.0$} & {1.82} & {$2.63 \pm 2.3^{*}$} & {1.90} \\ 
&\multirow{1}{*}{Homoscedastic}&Registration & $1.20 \pm 0.3$ & {1.20} & ${1.22} \pm {0.5}~$ & {1.07} & $2.05 \pm 1.0$ & 1.81 & $2.34 \pm 2.2$ & 1.60 \\ 
&\multirow{1}{*}{DWA}&Registration & {$1.22 \pm 0.3$} & 1.18 & {$1.37 \pm 0.7^{*}~$} & {1.20} & {$2.29 \pm 1.1^{*}$} & {2.04} & {$3.18 \pm 2.4^{*}$} & {2.43} \\ \hline 
 
\multirow{6}{*}{ Dense }&\multirow{2}{*}{Equal}&Segmentation & $1.14 \pm 0.4$ & 1.06 & \textcolor{light-gray}{$1.73 \pm 2.1~$} & \textcolor{light-gray}{1.12} & $1.91 \pm 0.9~$ & 1.64 & $1.04 \pm 0.7$ & 0.87 \\
&&Registration & \textcolor{light-gray}{$1.20 \pm 0.3$} & \textcolor{light-gray}{1.11} & $1.33 \pm 0.7^{*}~$ & 1.10 & \textcolor{light-gray}{$2.16 \pm 1.1$} & \textcolor{light-gray}{1.85} & \textcolor{light-gray}{$2.56 \pm 1.9$} & \textcolor{light-gray}{1.90} \\ [1mm]
&\multirow{2}{*}{Homoscedastic}&Segmentation & $1.09 \pm 0.3$ & 1.04 & \textcolor{light-gray}{$1.51 \pm 1.2~$} & \textcolor{light-gray}{1.13} & $1.86 \pm 0.8~$ & 1.69 & $0.99 \pm 0.4$ & 0.91 \\
&&Registration & \textcolor{light-gray}{$1.17 \pm 0.3$} & \textcolor{light-gray}{1.15} & $1.31 \pm 0.6~$ & 1.13 & \textcolor{light-gray}{$2.17 \pm 1.0$} & \textcolor{light-gray}{1.96} & \textcolor{light-gray}{$2.63 \pm 2.0^{*}$} & \textcolor{light-gray}{1.95} \\ [1mm]
&\multirow{2}{*}{DWA}&Segmentation & $1.12 \pm 0.3^{*\dagger}$ & 1.04 & \textcolor{light-gray}{$1.74 \pm 2.0~$} & \textcolor{light-gray}{1.13} & $1.99 \pm 0.9^{*}$ & 1.77 & $1.00 \pm 0.4~$ & 0.85 \\
&&Registration & \textcolor{light-gray}{$1.14 \pm 0.3$} & \textcolor{light-gray}{1.14} & $1.27 \pm 0.6~$ & {1.07} & \textcolor{light-gray}{$2.24 \pm 1.1^{*}$} & \textcolor{light-gray}{1.97} & \textcolor{light-gray}{$2.72 \pm 1.9$} & \textcolor{light-gray}{2.13} \\ \hline 
 
\multirow{6}{*}{ SEDD }&\multirow{2}{*}{Equal}&Segmentation & \textcolor{light-gray}{$1.47 \pm 0.6^{*\dagger}$} & \textcolor{light-gray}{1.31} & \textcolor{light-gray}{$2.81 \pm 4.6$} & \textcolor{light-gray}{1.34} & $1.97 \pm 1.0$ & 1.59 & $1.21 \pm 1.0$ & 0.94 \\
&&Registration & \textcolor{light-gray}{$1.28 \pm 0.4^{*}$} & \textcolor{light-gray}{1.19} & \textcolor{light-gray}{$1.50 \pm 0.9^{*}$} & \textcolor{light-gray}{1.26} & \textcolor{light-gray}{$2.26 \pm 1.1^{*}$} & \textcolor{light-gray}{1.94} & \textcolor{light-gray}{$2.61 \pm 2.1^{*}$} & \textcolor{light-gray}{1.83} \\ [1mm]
&\multirow{2}{*}{Homoscedastic}&Segmentation & $1.15 \pm 0.3^{\dagger}$ & 1.14 & \textcolor{light-gray}{$1.47 \pm 1.0$} & \textcolor{light-gray}{1.22} & $2.12 \pm 1.1$ & 1.91 & $0.99 \pm 0.2$ & 0.94 \\
&&Registration & \textcolor{light-gray}{$1.19 \pm 0.3$} & \textcolor{light-gray}{1.21} & $1.23 \pm 0.5~$ & 1.13 & \textcolor{light-gray}{$2.15 \pm 1.0$} & \textcolor{light-gray}{1.92} & \textcolor{light-gray}{$2.31 \pm 2.0$} & \textcolor{light-gray}{1.64} \\ [1mm]
&\multirow{2}{*}{DWA}&Segmentation & \textcolor{light-gray}{$1.22 \pm 0.3^{*\dagger}$} & 1.18 & \textcolor{light-gray}{$1.44 \pm 0.8$} & \textcolor{light-gray}{1.21} & $2.12 \pm 1.4$ & 1.73 & $1.10 \pm 0.6$ & 0.93 \\
&&Registration & $1.22 \pm 0.3$ & \textcolor{light-gray}{1.22} & $1.32 \pm 0.6^{*}~$ & 1.10 & \textcolor{light-gray}{$2.30 \pm 1.1^{*}$} & \textcolor{light-gray}{2.01} & \textcolor{light-gray}{$2.86 \pm 1.9^{*}$} & \textcolor{light-gray}{2.41} \\ \hline 
 
\multirow{6}{*}{Cross-stitch} & \multirow{2}{*}{Equal}&Segmentation & ${1.06} \pm {0.3}^{*}~$ & {0.99} & $1.27 \pm 0.4~$ & \textcolor{light-gray}{1.15} & ${1.76} \pm {0.8}^{*}~$ & {1.47} & ${0.91} \pm {0.4}~$ & {0.82} \\
&&Registration & \textcolor{light-gray}{$1.10 \pm 0.3^{*}~$} & \textcolor{light-gray}{1.06} & \textcolor{light-gray}{$1.30 \pm 0.6~$} & 1.13 & \textcolor{light-gray}{$2.00 \pm 1.0^{*}~$} & \textcolor{light-gray}{1.75} & \textcolor{light-gray}{$2.45 \pm 2.1~$} & \textcolor{light-gray}{1.81} \\ [1mm]
&\multirow{2}{*}{Homoscedastic}&Segmentation & \textcolor{light-gray}{$1.23 \pm 0.3^{\dagger}$} & \textcolor{light-gray}{1.16} & \textcolor{light-gray}{$1.51 \pm 1.2~$} & \textcolor{light-gray}{1.17} & \textcolor{light-gray}{$2.37 \pm 1.0$} & \textcolor{light-gray}{2.09} & $0.92 \pm 0.2~$ & 0.89 \\
&&Registration & \textcolor{light-gray}{$1.24 \pm 0.3$} & \textcolor{light-gray}{1.24} & $1.32 \pm 0.6~$ & 1.13 & $2.12 \pm 1.0$ & 1.89 & \textcolor{light-gray}{$2.45 \pm 1.9$} & \textcolor{light-gray}{1.97} \\ [1mm]
&\multirow{2}{*}{DWA}&Segmentation & \textcolor{light-gray}{$1.34 \pm 0.4^{*\dagger}$} & \textcolor{light-gray}{1.27} & \textcolor{light-gray}{$1.75 \pm 1.7$} & \textcolor{light-gray}{1.29} & \textcolor{light-gray}{$2.32 \pm 0.9^{\dagger}$} & \textcolor{light-gray}{2.11} & $1.17 \pm 0.8^{*}$ & 0.91 \\
&&Registration & $1.22 \pm 0.3$ & 1.19 & $1.27 \pm 0.6~$ & 1.09 & $2.21 \pm 1.0^{*}~$ & 2.00 & \textcolor{light-gray}{$2.93 \pm 2.3^{*}$} & \textcolor{light-gray}{2.27} \\ \hline 
 
		\end{tabular}
	}
	\label{table:HMC_MSD_weighting}
\end{table*}

\begin{figure*}[t!]
\begin{center}
\includegraphics[width=1\textwidth]{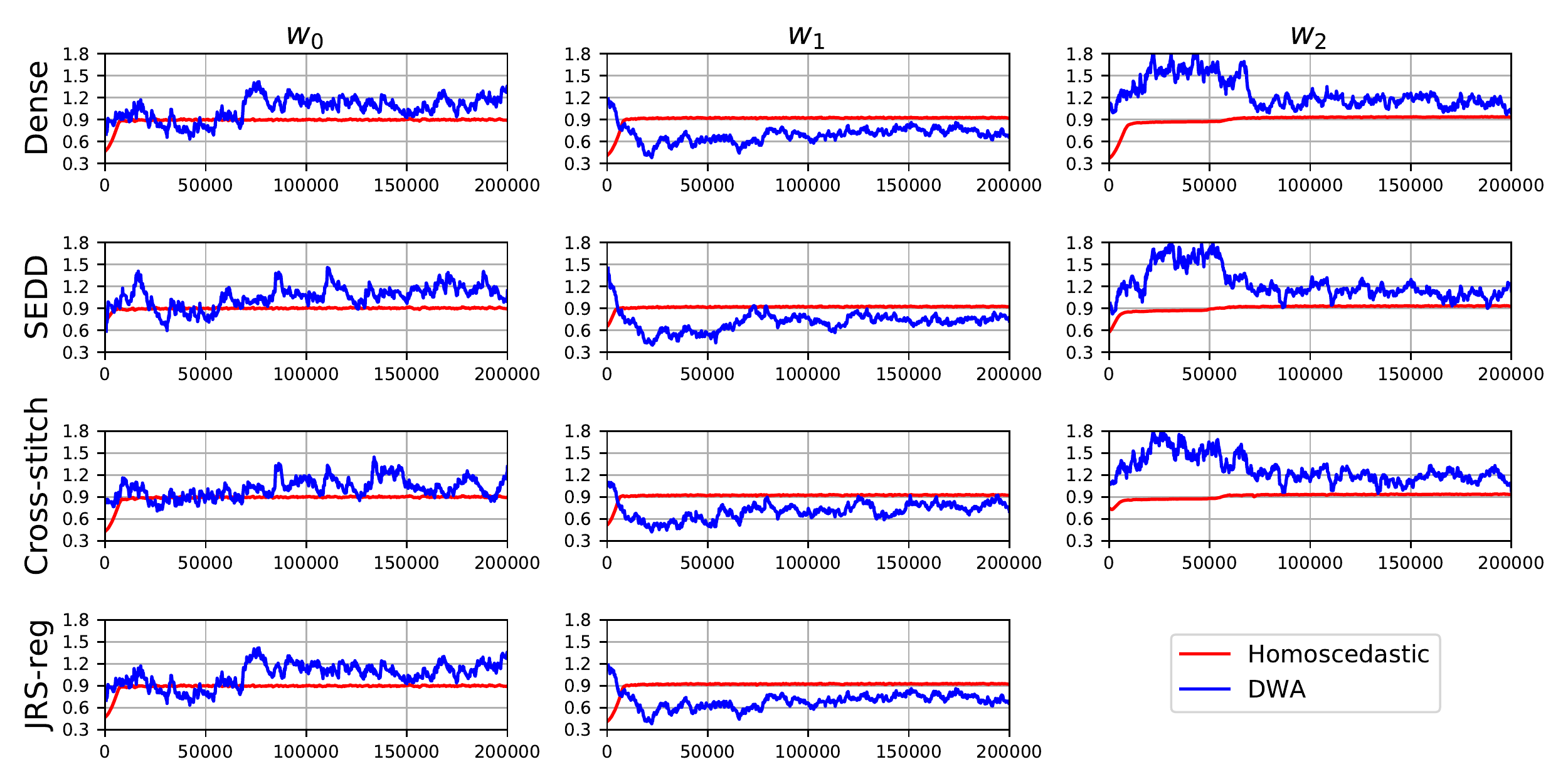}
\caption{The evolution of the loss weights during training for different multi-task networks on the validation dataset (HMC).}
\label{fig:loss-weights}
\end{center}
\end{figure*}

\begin{figure}[t]
\begin{center}
\includegraphics[width=1\linewidth]{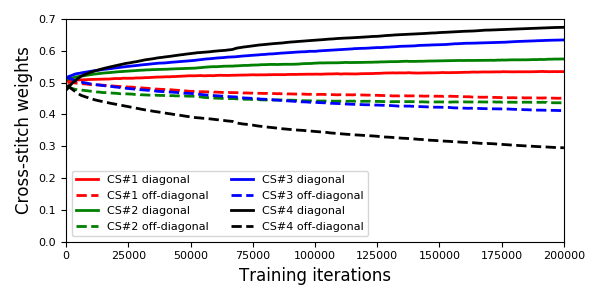}
\caption{The evolution of the Cross-stitch units weights during training using equal weights. CS\#1 and CS\#2 are placed in the down-sampling path, while CS\#3 and CS\#4 are placed in the upsampling path. The solid lines represent the mean of the weights across the diagonal of the CS unit, while the dashed lines represent the mean of the off-diagonal weights. }
\label{fig:CS-weights}
\end{center}
\end{figure}

\begin{figure}[t]
\begin{center}
\includegraphics[width=1\linewidth]{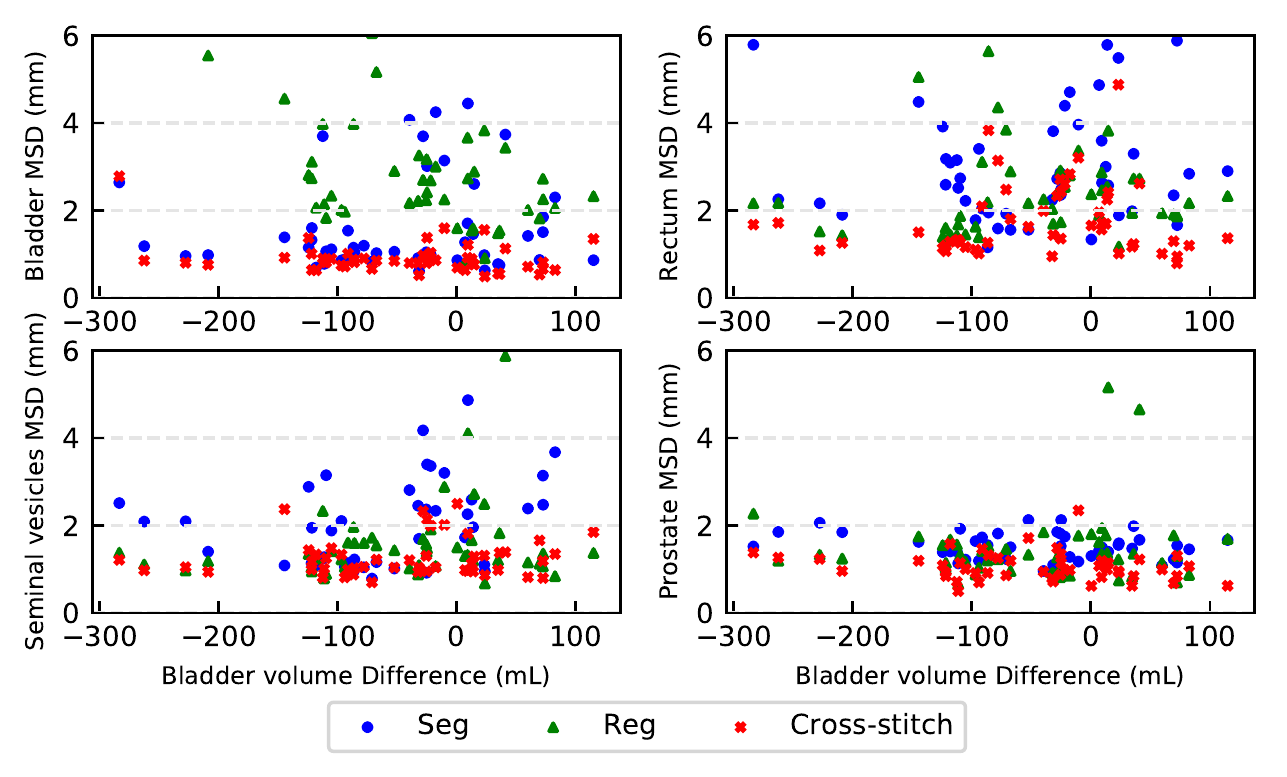}
\caption{The effect of the bladder volume deviation from the planning volume on the performance of the Seg, Reg, and Cross-stitch networks for the validation set (HMC). }
\label{fig:HMC_bladder_filling}
\end{center}
\end{figure}

\begin{figure}[t]
\begin{center}
\includegraphics[width=1\linewidth]{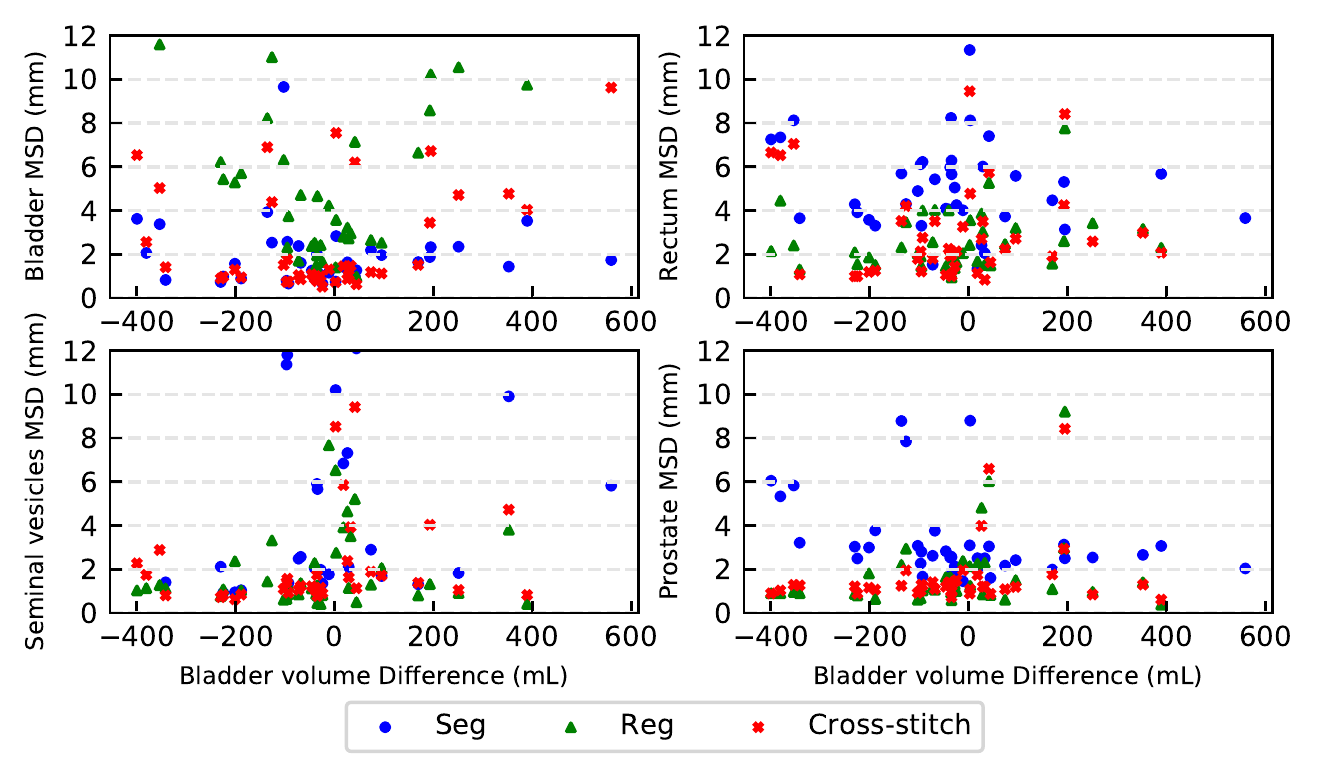}
\caption{The effect of the bladder volume deviation from the planning volume on the performance of the STL and the Seg, Reg, and Cross-stitch networks for the independent test set (EMC). }
\label{fig:EMC_bladder_filling}
\end{center}
\end{figure}

\subsection{Optimization of the Networks Inputs}
During training, validation, and testing, we have access to the fixed image \IF, the moving image \IM, and the moving segmentation \SM. In Table \ref{table:network_input_MSD} we compared different sets of inputs on the validation dataset. This experiment helps to better understand how these network interpret and utilize these inputs and how this would reflect on the network outcome represented by the MSD metric. For this experiment we used equal loss weights for the MTL networks. 

Feeding \SM to the segmentation network improves the results substantially compared to only feeding \IF, especially for the seminal vesicles, while feeding \IM deteriorates the results. 
For the registration and JRS-reg networks, feeding \SM alongside \IF and \IM resulted in a similar performance compared to not feeding it.
Since the Cross-stitch network is composed of two networks, one for segmentation and the other for registration, we experimented with various combinations of inputs. The results are very consistent with our previous findings on the single-task networks on the effect of using \SM as an input.

For the remainder of this paper, we chose to use \IF as input for the segmentation network, and \IF and \IM as inputs for the registration network. Although adding \SM proved to be better especially for the segmentation network, here we exclude it, since these two methods act as a baseline and this is the standard setting in single-task networks. For dense, SEDD, and JRS-reg networks, we select a concatenation of \IM, \IF, and \SM for the final network. For the Cross-stitch network, we select \IF for the segmentation network and the concatenation of \IM, \IF, and \SM for the registration network. 

\subsection{Optimization of loss weighting strategy}

In this experiment we investigate the performance of the various loss weighting strategies introduced in Section \ref{loss_weighting} in order to select the best weighting method for the underlying tasks. 

Table \ref{table:HMC_MSD_weighting} shows the results of the different weighting strategies for the MTL networks in terms of MSD. 
For the JRS-reg network architecture, weighting the losses with homoscedastic uncertainty achieved comparable results to using equal weights, while DWA scored somewhat less. For the dense and SEDD architectures, homoscedastic weighting achieved a slightly better performance, while equal weights was best for the Cross-stitch network. For these architectures (dense, SEDD, and Cross-stitch), the segmentation output path showed improvement over the registration output path.

Figure \ref{fig:loss-weights} illustrates the evolution of the loss weights $w_i$ during training, for different multi-task network architectures and weighting strategies. 

For the remainder of this paper and based on the previous findings, we chose the homoscedastic uncertainty weighting strategy for the JRS-reg, dense and SEDD networks, while using equal weights for the Cross-stitch network. 


\subsection{Analysis of Cross-stitch units}
Analysis of the behavior of the Cross-stitch units during training facilitates the understanding of how the segmentation and registration networks interacts in the MTL settings. Figure \ref{fig:CS-weights} shows the mean of the CS units across the diagonal and off-diagonal (See Equation (\ref{eq:cs})). Higher weights on the diagonal means that the network tends to separate the task-specific feature maps, while higher weights off-diagonal means that the network tends to share the corresponding feature maps.

\subsection{Effect of the bladder filling}
For the HMC dataset,  which was used for training and validation, a bladder filling protocol was in place, meaning that the deformation of the bladder between daily and planning scans is not large. However, this is not the scenario for the EMC dataset, the test set. 

Figure \ref{fig:HMC_bladder_filling} and \ref{fig:EMC_bladder_filling} illustrates the effect of the bladder volume variation from the planning scan on the performance of the Seg, Reg, and Cross-stitch networks. The Cross-stitch network is resilient to bladder filling for both the HMC and EMC datasets.  

\subsection{Evaluation of the Quality of the DVF}

The smoothness of the predicted DVF is an important parameter to evaluate the predicted deformation field. Table \ref{table:dvf-table} shows a detailed analysis of the DVF in terms of the standard deviation of the determinant of the Jacobian as well as the folding fraction for the registration path of the different networks.

\begin{table*}[!htb]
	\centering
	\setlength{\tabcolsep}{3pt}
	\caption[]{MSD (mm) values for the different networks on the validation set (HMC). Lower values are better.}
	\resizebox{\textwidth}{!}{
		\begin{tabular}{lllclclclc} 
			&&\multicolumn{2}{c}{Prostate}&\multicolumn{2}{c}{Seminal vesicles}&\multicolumn{2}{c}{Rectum}& \multicolumn{2}{c}{Bladder} \\ \hline
			Network & Output path & \multicolumn{1}{c}{$\mu \pm \sigma$} & median & \multicolumn{1}{c}{$\mu \pm \sigma$} & median & \multicolumn{1}{c}{$\mu \pm \sigma$} & median & \multicolumn{1}{c}{$\mu \pm \sigma$} & median \\ \hline
\multirow{1}{*}{ Seg }&Segmentation & $1.49 \pm 0.3$ & 1.49 & $2.50 \pm 2.6$ & 2.09 & $3.39 \pm 2.2$ & 2.73 & $1.60 \pm 1.1$ & 1.13 \\ \hline 
 
\multirow{1}{*}{ Reg }&Registration & $1.43 \pm 0.8$ & 1.29 & $1.71 \pm 1.4$ & 1.37 & $2.44 \pm 1.1$ & 2.17 & $3.40 \pm 2.3$ & 2.71 \\ \hline

\multirow{1}{*}{ JRS-reg }&Registration & $1.20 \pm 0.3$ & \textcolor{light-gray}{1.20} & ${1.22} \pm {0.5}~$ & {1.07} & $2.05 \pm 1.0$ & 1.81 & $2.34 \pm 2.2$ & 1.60 \\ \hline

\multirow{2}{*}{ Dense }&Segmentation & $1.09 \pm 0.3$ & 1.04 & \textcolor{light-gray}{$1.51 \pm 1.2~$} & \textcolor{light-gray}{1.13} & $1.86 \pm 0.8~$ & 1.69 & $0.99 \pm 0.4$ & 0.91 \\
&Registration & \textcolor{light-gray}{$1.17 \pm 0.3$} & \textcolor{light-gray}{1.15} & $1.31 \pm 0.6~$ & 1.13 & \textcolor{light-gray}{$2.17 \pm 1.0$} & \textcolor{light-gray}{1.96} & \textcolor{light-gray}{$2.63 \pm 2.0$} & \textcolor{light-gray}{1.95} \\ \hline

\multirow{2}{*}{ SEDD }&Segmentation & $1.15 \pm 0.3$ & 1.14 & \textcolor{light-gray}{$1.47 \pm 1.0$} & \textcolor{light-gray}{1.22} & $2.12 \pm 1.1$ & 1.91 & $0.99 \pm 0.2$ & 0.94 \\
&Registration & \textcolor{light-gray}{$1.19 \pm 0.3$} & \textcolor{light-gray}{1.21} & $1.23 \pm 0.5~$ & 1.13 & \textcolor{light-gray}{$2.15 \pm 1.0$} & \textcolor{light-gray}{1.92} & \textcolor{light-gray}{$2.31 \pm 2.0$} & \textcolor{light-gray}{1.64} \\ \hline 
 
\multirow{2}{*}{ Cross-stitch }&Segmentation & ${1.06} \pm {0.3}$ & {0.99} & $1.27 \pm 0.4~$ & \textcolor{light-gray}{1.15} & ${1.76} \pm {0.8}$ & {1.47} & ${0.91} \pm {0.4}~$ & {0.82} \\
&Registration & \textcolor{light-gray}{$1.10 \pm 0.3$} & \textcolor{light-gray}{1.06} & \textcolor{light-gray}{$1.30 \pm 0.6~$} & 1.13 & \textcolor{light-gray}{$2.00 \pm 1.0$} & \textcolor{light-gray}{1.75} & \textcolor{light-gray}{$2.45 \pm 2.1~$} & \textcolor{light-gray}{1.81} \\ \hline 
 
\multirow{1}{*}{ Elastix \cite{qiao2017fast}}&Registration & $1.73 \pm 0.7$ & 1.59 & $2.71 \pm 1.6$ & 2.45 & $3.69 \pm 1.2$ & 3.50 & $5.26 \pm 2.6$ & 4.72 \\ \hline 
 
\multirow{1}{*}{ Hybrid \cite{MedPhys}}&Registration & $1.27 \pm 0.3$ & 1.25 & $1.47 \pm 0.5$ & 1.32 & $2.03 \pm 0.6$ & 1.85 & $1.75 \pm 1.0$ & 1.26 \\ \hline

\multirow{1}{*}{ JRS-GAN \cite{JrsGan}}&Registration & $1.14 \pm 0.3$ & 1.04 & $1.75 \pm 1.3$ & 1.44 & $2.17 \pm 1.1$ & 1.89 & $2.25 \pm 1.9$ & 1.54 \\ \hline 
 
		\end{tabular}
	}
	\label{table:HMC_MSD}
\end{table*}


\begin{table*}[!htb]
	\centering
	\setlength{\tabcolsep}{3pt}
	\caption[Table caption text]{MSD (mm) values for the different networks on the independent test set (EMC). Lower values are better. }
	\resizebox{\textwidth}{!}{
		\begin{tabular}{lllclclclc} 
			&&\multicolumn{2}{c}{Prostate}&\multicolumn{2}{c}{Seminal vesicles}&\multicolumn{2}{c}{Rectum}& \multicolumn{2}{c}{Bladder} \\ \hline
			Network & Output path & \multicolumn{1}{c}{$\mu \pm \sigma$} & median & \multicolumn{1}{c}{$\mu \pm \sigma$} & median & \multicolumn{1}{c}{$\mu \pm \sigma$} & median & \multicolumn{1}{c}{$\mu \pm \sigma$} & median \\ \hline
\multirow{1}{*}{ Seg }&Segmentation & $3.18 \pm 1.8$ & 2.57 & $9.33 \pm 10.1$ & 5.82 & $5.79 \pm 3.4$ & 5.18 & $1.88 \pm 1.5~$ & 1.50 \\ \hline 
 
\multirow{1}{*}{ Reg }&Registration & $2.01 \pm 2.5$ & 1.18 & $2.86 \pm 5.2$ & 1.18 & $2.89 \pm 2.5$ & 2.23 & $5.98 \pm 4.7$ & 4.44 \\ \hline

\multirow{1}{*}{ JRS-reg }&Registration & $1.94 \pm 2.6$ & 1.16 & $2.48 \pm 4.8~$ & 1.01 & \textcolor{light-gray}{$2.67 \pm 2.4$} & 2.05 & $4.80 \pm 4.6$ & 2.12 \\ \hline

\multirow{2}{*}{ Dense }&Segmentation & \textcolor{light-gray}{$2.01 \pm 2.6$} & 1.15 & \textcolor{light-gray}{$4.08 \pm 7.2$} & \textcolor{light-gray}{1.23} & \textcolor{light-gray}{$3.70 \pm 5.4~$} & 2.03 & $2.75 \pm 3.1$ & 1.23 \\
&Registration & $1.93 \pm 2.5$ & \textcolor{light-gray}{1.15} & $2.53 \pm 4.7~$ & 1.01 & $2.67 \pm 2.3$ & \textcolor{light-gray}{2.13} & \textcolor{light-gray}{$5.08 \pm 4.4$} & \textcolor{light-gray}{3.01} \\ \hline

\multirow{2}{*}{ SEDD }&Segmentation & \textcolor{light-gray}{$1.99 \pm 2.4$} & \textcolor{light-gray}{1.24} & \textcolor{light-gray}{$6.26 \pm 8.9$} & \textcolor{light-gray}{3.01} & \textcolor{light-gray}{$4.21 \pm 4.9~$} & 2.12 & $2.43 \pm 2.9$ & 1.04 \\
&Registration & $1.92 \pm 2.5$ & 1.19 & $2.43 \pm 4.5~$ & 1.07 & $2.72 \pm 2.4$ & \textcolor{light-gray}{2.17} & \textcolor{light-gray}{$4.86 \pm 4.4$} & \textcolor{light-gray}{2.22} \\ \hline

\multirow{2}{*}{ Cross-stitch }&Segmentation & $1.88 \pm 1.9~$ & \textcolor{light-gray}{1.30} & \textcolor{light-gray}{$2.76 \pm 3.5~$} & \textcolor{light-gray}{1.28} & \textcolor{light-gray}{$4.87 \pm 6.8~$} & \textcolor{light-gray}{2.49} & $1.66 \pm 1.7$ & 0.85 \\
&Registration & \textcolor{light-gray}{$1.91 \pm 2.3$} & 1.23 & $2.41 \pm 4.5~$ & {0.95} & $2.78 \pm 2.4$ & 2.16 & \textcolor{light-gray}{$4.90 \pm 4.0$} & \textcolor{light-gray}{2.84} \\ \hline

\multirow{1}{*}{ Elastix~\cite{qiao2017fast} }&Registration & ${1.42} \pm {0.7}~$ & 1.17 & $2.07 \pm 2.6$ & 1.24 & $3.20 \pm 1.6$ & 3.07 & $5.30 \pm 5.1$ & 3.27 \\ \hline 
 
\multirow{1}{*}{ Hybrid~\cite{MedPhys} }&Registration & $1.55 \pm 0.6$ & 1.36 & ${1.65} \pm {1.3}~$ & 1.22 & $2.65 \pm 1.6~$ & 2.36 & $3.81 \pm 3.6$ & 2.26 \\ \hline

		\end{tabular}
	}
	\label{table:EMC_MSD}
\end{table*}

\begin{table*}[h]
\centering
\caption{Analysis of the determinant of the Jacobian for the validation and the independent test sets. Lower values are better. }
\label{table:dvf-table}
\begin{tabular}{lcccccc}
&&\multicolumn{2}{c}{Validation set (HMC) }&&\multicolumn{2}{c}{Independent test set (EMC) } \\
\hline

Network && Std. Jacobian & Folding fraction && Std. Jacobian & Folding fraction
\\ \hline
Reg &&$0.2935\pm0.1022$ & $0.0049\pm0.0039$ && $0.4129\pm0.2258$ & $0.0112\pm0.0115$ \\ \hline

JRS-reg && $0.2543\pm0.0505$ & $0.0030\pm0.0014$ && $0.3148\pm0.1106$ & $0.0066\pm0.0062$ \\ \hline

Dense && $0.2062\pm0.0431$ & $0.0018\pm0.0012$ && 
$0.2558\pm0.0899$ & $0.0036\pm0.0027$ \\ \hline

SEDD && $0.2626\pm0.1167$ & $0.0019\pm0.0016$ && 
$0.4287\pm0.3000$ & $0.0066\pm0.0074$ \\ \hline

Cross-stitch && $0.2241\pm0.0784$ & $0.0024\pm0.0018$ && $0.3301\pm0.1869$ &  $0.0071\pm0.0070$ \\ \hline

\end{tabular}
\end{table*}

\subsection{Comparison against the state-of-the-art}

Table \ref{table:HMC_MSD} and \ref{table:EMC_MSD} show the results for the validation set (HMC) and test set (EMC), respectively. The first two networks in each table are single-task networks. For both sets, the registration network outperformed the segmentation network for all organs except the bladder. The mean MSD for the independent test set is higher than the corresponding numbers in the validation set for most organs. However, the median values are on par. For the MTL networks, the segmentation path of the networks achieved better performance than the registration path on both datasets except for the seminal vesicles. The Cross-stitch network achieved the best results compared to the other MTL networks. 

The proposed STL and MTL networks were compared against other state-of-the-art methods that were evaluated using the HMC dataset. For the validation set, the STL network achieved comparable results, while the Cross-stitch network outperformed these methods for both output paths. On the test set, \textit{elastix} \cite{qiao2017fast} and the Hybrid method \cite{MedPhys} performed better except for the bladder, although the median values of the MTL networks were better. 

\begin{figure*}[t]
	\centering
	\resizebox{\textwidth}{!}{
		\begin{tabular}{c @{\quad} c @{\quad} || c @{\quad} c @{\quad} || c @{\quad} c}
			
			 \large{Seg} & \large{Reg} & \large{Seg} & \large{Reg} & \large{Seg} & \large{Reg} \\
			
			\includegraphics[width=25mm,height=35mm]{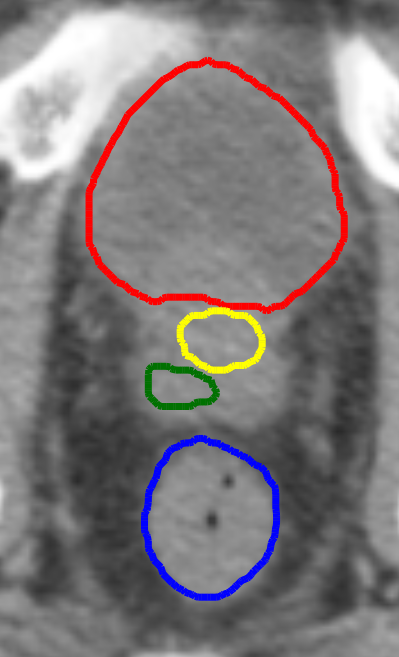} &
			\includegraphics[width=25mm,height=35mm]{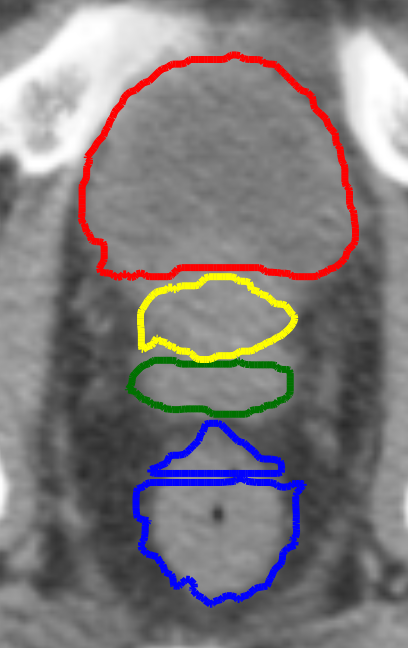} &
			\includegraphics[width=25mm,height=35mm]{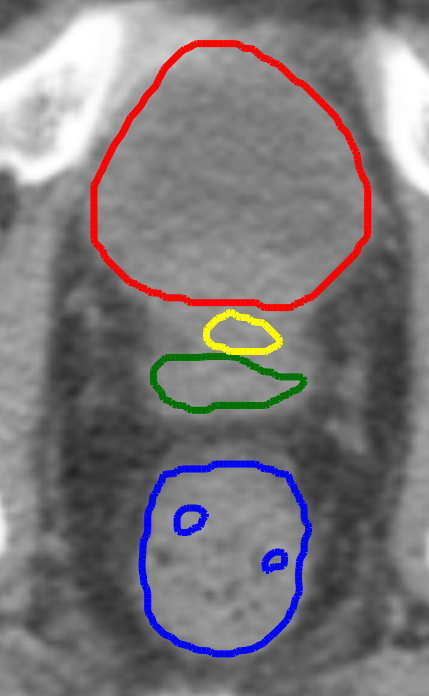} &
			\includegraphics[width=25mm,height=35mm]{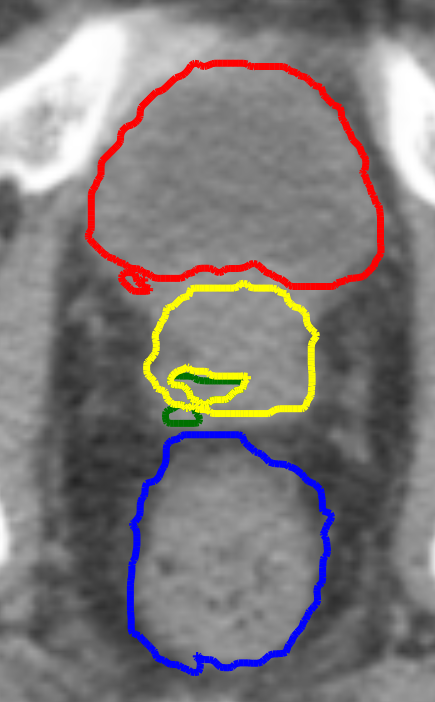} &
			\includegraphics[width=25mm,height=35mm]{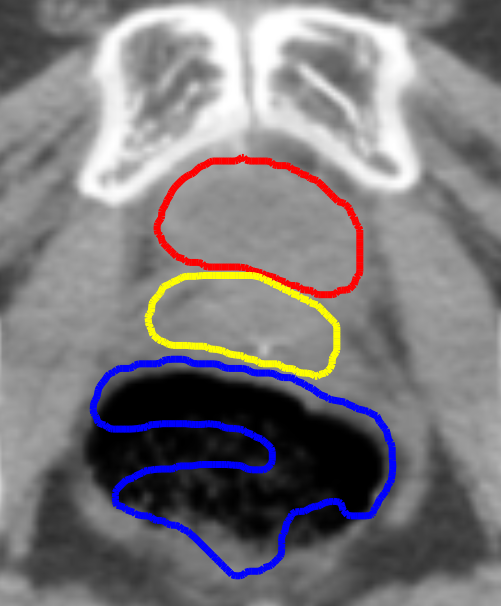} &
			\includegraphics[width=25mm,height=35mm]{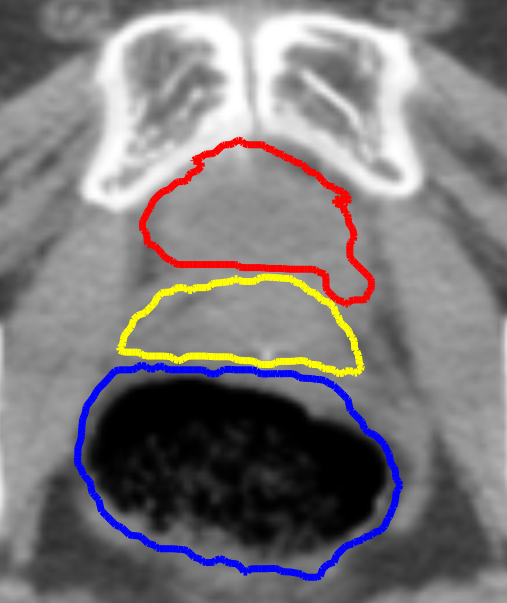} \\
			
			 \large{Cross-stitch} & \large{Manual} & \large{Cross-stitch} & \large{Manual} & \large{Cross-stitch} & \large{Manual} \\
			
            \includegraphics[width=25mm,height=35mm]{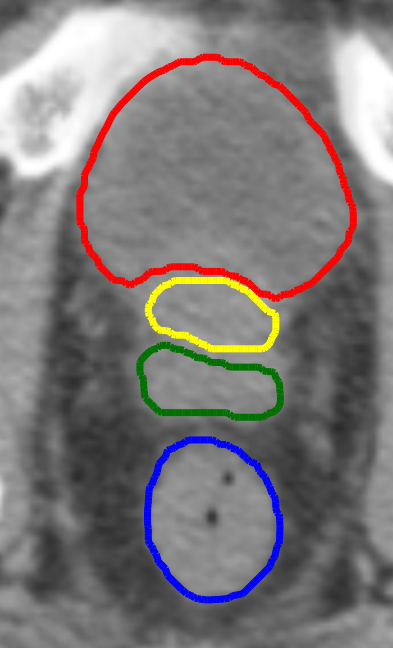} &
			\includegraphics[width=25mm,height=35mm]{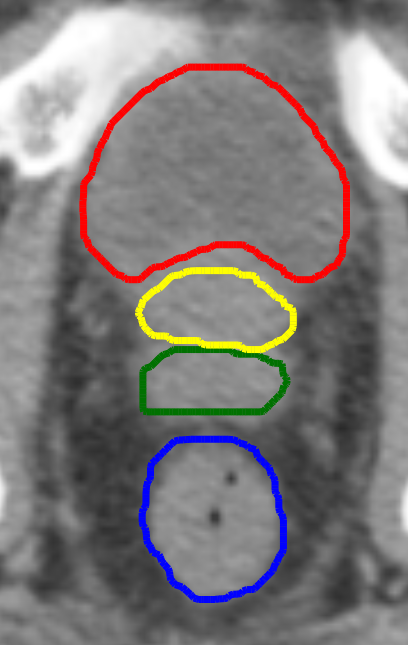} &
			\includegraphics[width=25mm,height=35mm]{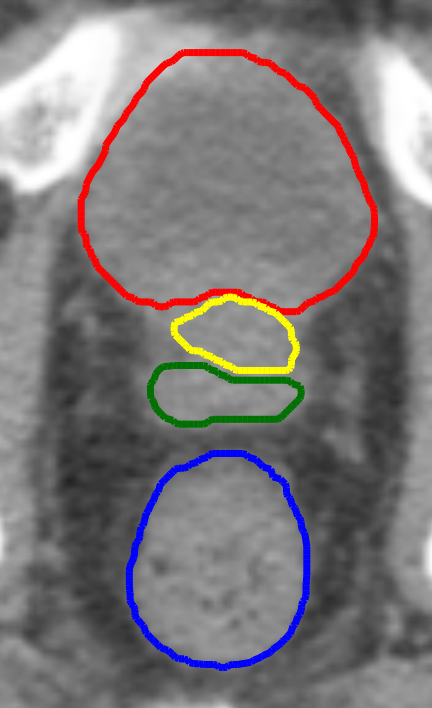} &
			\includegraphics[width=25mm,height=35mm]{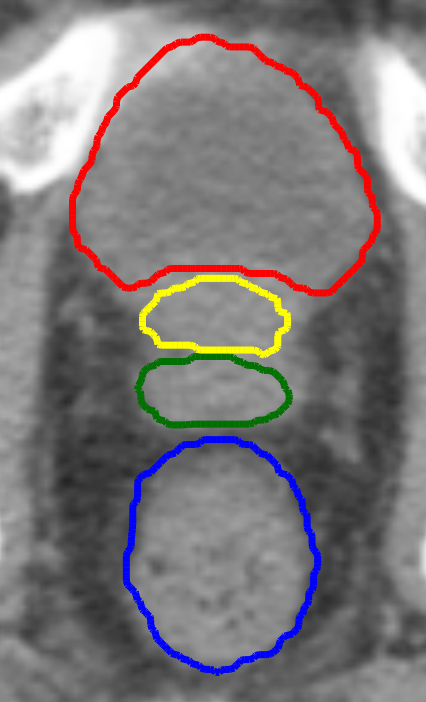} &
			\includegraphics[width=25mm,height=35mm]{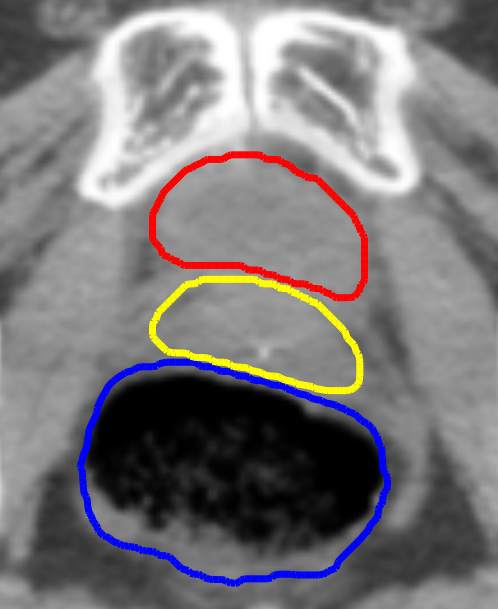} &
			\includegraphics[width=25mm,height=35mm]{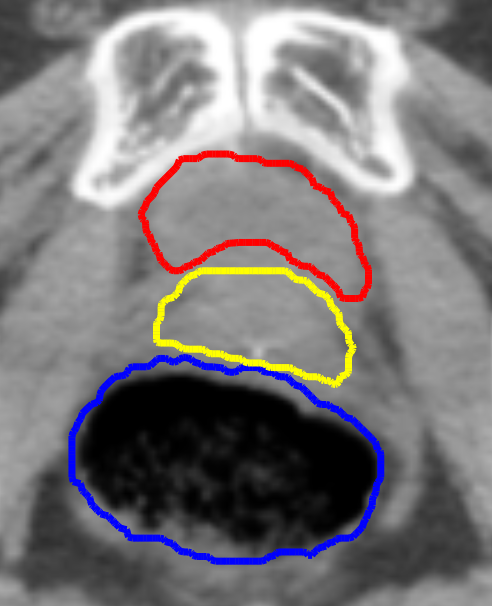} 
			
		\end{tabular}
}
\caption{Example contours from the validation dataset (HMC) generated by the proposed STL and MTL networks. From left to right, the selected cases are the first, second, and third quartile in terms of the prostate MSD of the Cross-stitch network. The contours of the bladder, prostate, seminal vesicles, and rectum are colored in red, yellow, green, and blue, respectively.}
\label{fig:HMC_examples}
\end{figure*}

\begin{figure*}[t]
	\centering
	\resizebox{\textwidth}{!}{
		\begin{tabular}{c @{\quad} c @{\quad} || c @{\quad} c @{\quad} || c @{\quad} c}
			
			 \large{Seg} & \large{Reg} & \large{Seg} & \large{Reg} & \large{Seg} & \large{Reg} \\
			
			\includegraphics[width=25mm,height=35mm]{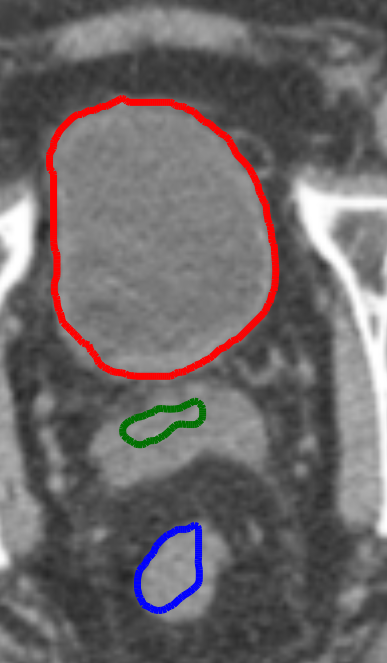} &
			\includegraphics[width=25mm,height=35mm]{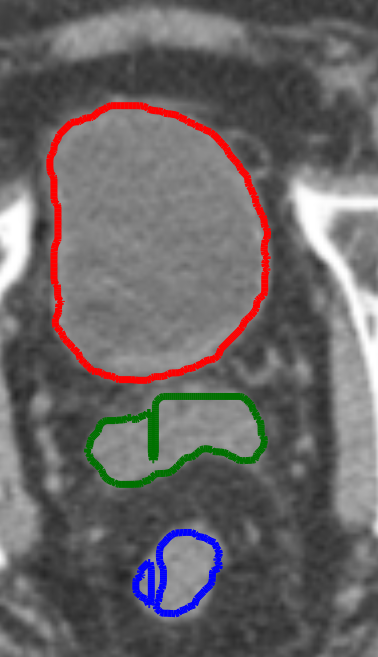} &
			\includegraphics[width=25mm,height=35mm]{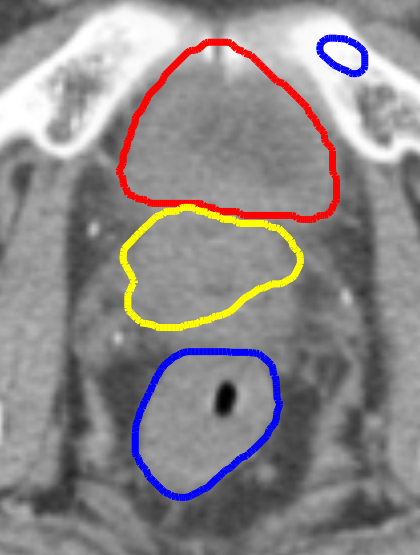} &
			\includegraphics[width=25mm,height=35mm]{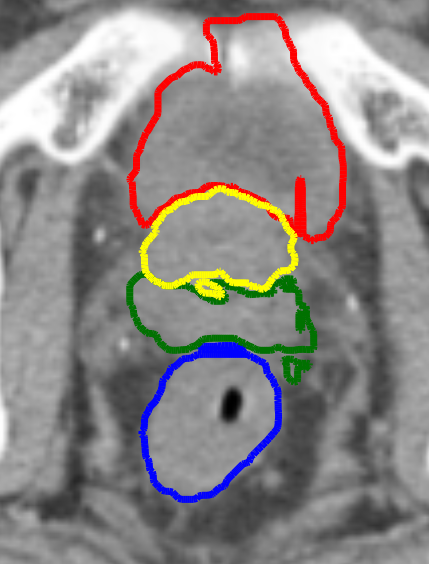} &
			\includegraphics[width=25mm,height=35mm]{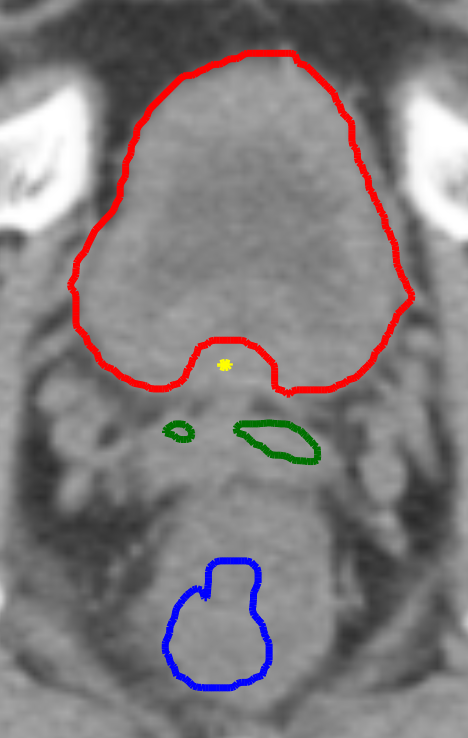} &
			\includegraphics[width=25mm,height=35mm]{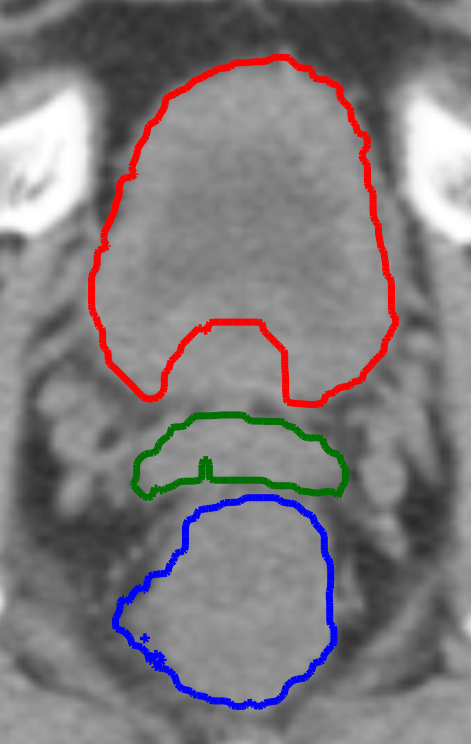} \\
			
			 \large{Cross-stitch} & \large{Manual} & \large{Cross-stitch} & \large{Manual} & \large{Cross-stitch} & \large{Manual}\\
			
            \includegraphics[width=25mm,height=35mm]{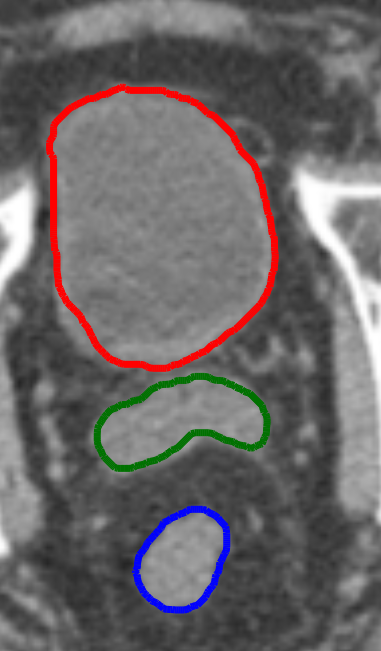} &
			\includegraphics[width=25mm,height=35mm]{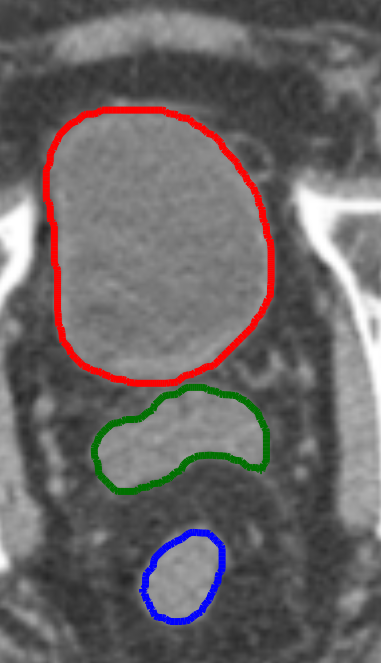} &
			\includegraphics[width=25mm,height=35mm]{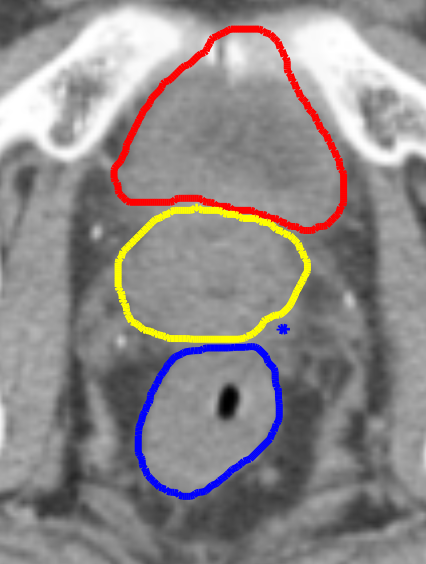} &
			\includegraphics[width=25mm,height=35mm]{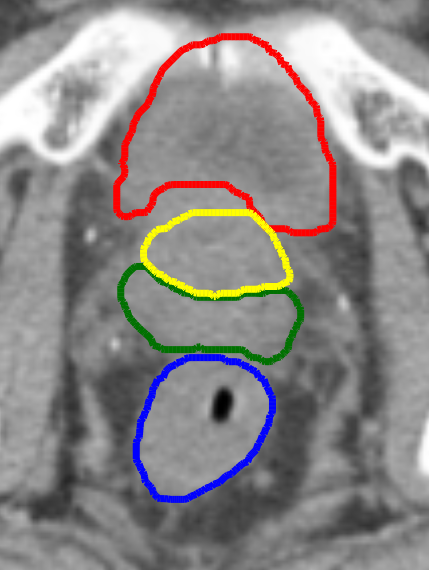} &
			\includegraphics[width=25mm,height=35mm]{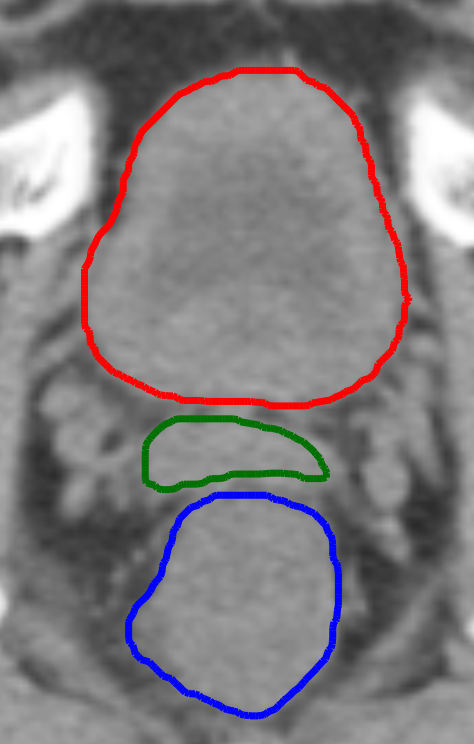} &
			\includegraphics[width=25mm,height=35mm]{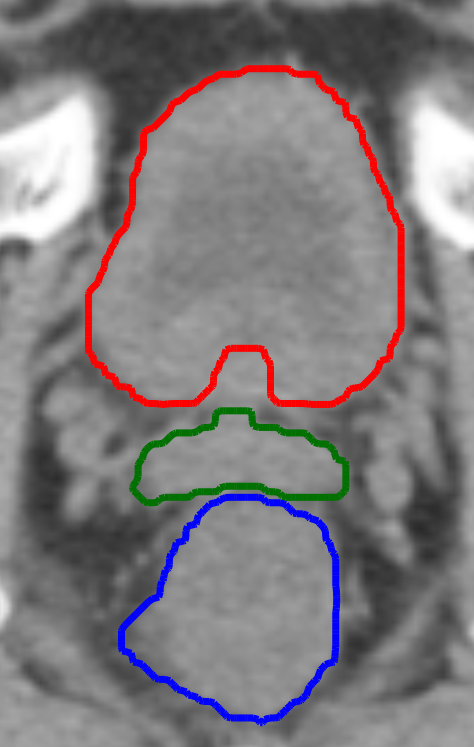} 
			
		\end{tabular}
		}
\caption{Example contours from the independent test set (EMC) generated by the proposed STL and MTL networks. From left to right, the selected cases are the first, second, and third quartile in terms of the prostate MSD of the Cross-stitch network.}
\label{fig:EMC_examples}
\end{figure*}

For the quality of the predicted contours, Figure \ref{fig:HMC_examples} and \ref{fig:EMC_examples} show example contours from the HMC and EMC datasets for the Seg, Reg, and Cross-stitch networks. The examples show that the Cross-stitch network achieves better results compared to the Seg and Reg networks especially for the seminal vesicles and rectum with large gas pockets.

\section{Discussion} \label{discussion}
In this study, we proposed to merge image registration and segmentation on the architectural level as well as via the loss, via a multi-task learning setting. We studied different network architectures and loss weighting methods in order to explore how these tasks interact, and thereby leverage the shared knowledge between them. Moreover, we carried out extensive quantitative analysis in the context of adaptive radiotherapy, and compared the proposed multi-task methods to their single-task counterparts. In this paper, a substantial number of experiments were executed, where we explored the following methodological choices: the bending energy weight, the input to the STL and MTL networks, and the loss weighting method. We also performed a thorough analysis on how Cross-stitch units and loss weights evolve during training. Finally, we compared our proposed methods against state-of-the-art methods.

In all the experiments we fixed the weight of the bending energy weight so that the network would not set it too low in order to improve the DSC of the deformed contours on the account of the smoothness of the predicted DVF. As shown in Figure \ref{fig:bending_energy} low bending energy weights result in better contour quality on the account of the smoothness of the predicted DVF.

For the inputs to the STL networks, additionally feeding \SM to the segmentation network resulted in a statistically significant improvement especially for the seminal vesicles. Apparently the network considers \SM as an initial estimation for \SF and subsequently uses it as a guidance for its final prediction. When feeding \IM the results deteriorated; this may confuse the network as \IF and \IM have the same anatomy but with different shapes and local positions. The addition of both \IM and \SM performed similar to the addition of only \SM, which indicates that the networks learned to ignore \IM. For the registration network, the addition of \SM resulted in a sub-optimal result, since the \SM contours on its own does not represent the underlying deformation well.

For the inputs to the MTL networks, in the JRS-reg network, feeding \SM alongside \IF and \IM resulted in a similar performance compared to not feeding it. This indicates that the incorporation of \SM via the DSC loss, already enables the JRS-reg network to exploit this extra information, and that additionally adding \SM as a network input does not provide further benefits. In the Cross-stitch network, we found that adding \SM to the registration network results in a statistically significant improvement. Furthermore, feeding \SM to one of the networks is sufficient, proving that segmentation and registration networks communicate their knowledge efficiently through the Cross-stitch units.

We selected the STL networks with \IF (for segmentation) and \IF alongside \IM (for registration) as input to our baseline methods. Between these two networks, the registration network performed better overall, since the registration network leverages prior knowledge from the organs in the moving image. For the bladder, the segmentation network achieved better results; Apparently the registration network had difficulties finding the correspondence between the bladder in the fixed and moving images, since it tends to deform considerably between visits. However, the segmentation network failed to segment the seminal vesicles for five cases. That is explained by the fact that the seminal vesicles is a difficult structure to segment, due to its relatively small size, undefined borders, and poor contrast with its surroundings. The registration network on the other hand is able to employ the surrounding anatomy as context, to accurately warp the seminal vesicles. 

For the multi-task networks, we demonstrated that fusing segmentation and registration tasks is performing better than its single-task counterparts. Merging these tasks using Cross-stitch network achieved the best results on both the validation and testing datasets.



Different loss weighting methods achieved comparable results as shown in Table \ref{table:HMC_MSD_weighting}. In Figure \ref{fig:loss-weights}, homoscedastic uncertainty tended to weigh all losses equally, using almost a fixed weight of 0.9 during most of the training iterations. On the contrary, DWA tended to fluctuate during training as the weights are updated based on the ratio of the loss from previous iterations, which fluctuates due to the batch-based training. Since the fixed and moving images are affinely registered beforehand, DWA tended to down-weigh the registration loss and the associated DSC at the beginning of the training, while weighting the segmentation network loss more in order to improve its prediction. Later during training, all the weights stabilized around 0.9 similar to homoscedastic uncertainty. Although both methods stabilized by the end of the training around the same value (0.9), the homoscedastic uncertainty achieved slightly better results compared to DWA and equal weighting methods, except for the Cross-stitch network. Our reasoning behind this is that homoscedastic uncertainty, unlike other methods, is learnable during the training and highly dependent on the underlying task uncertainty.



By analyzing the performance of the Cross-stitch units as demonstrated in Figure \ref{fig:CS-weights}, we found that the Cross-stitch units tended to average feature maps for the down-sampling path, while preferring to be more task-specific for the upsampling path. This somewhat mimics the shared encoder double decoder (SEDD) network, but in contrast to this network, the Cross-stitch network does not completely split the decoder paths. This finding confirms that the segmentation and registration tasks are correlated and thereby encode similar features.

We carried out an experiment to study the effect of the bladder filling protocol between the HMC and EMC datasets. As shown in Figure \ref{fig:HMC_bladder_filling}, the HMC dataset has a bladder filling protocol so the volume of the bladder changes slightly around 100 mL between different sessions, which is not the case for the EMC dataset as shown in Figure \ref{fig:EMC_bladder_filling}. Since the registration-based networks and joint networks were trained on small bladder  deformations, they failed on large  deformations, however the segmentation network was not affected since it does not depend on the deformation but rather the underlying texture to segment the bladder. 

In terms of the smoothness of the predicted DVF shown in Table \ref{table:dvf-table}, MTL networks achieved lower numbers for the standard deviation of the Jacobian as well as for the folding fraction, compared to the STL network (Reg), on both the test and validation set. Our reasoning is that joining the segmentation task to the registration task works as an additional regularization to the registration network. Due to the fact that the higher the quality of the predicted DVF, the higher the quality of the propagated contours and subsequently the lower the DSC loss. The numbers on the test set are slightly higher than the validation set, but this is due to the variance between the deformations between both sets and the fact that the network has not seen the test set before. This can be addressed using transfer learning as suggested by Elmahdy \textit{et al.} \cite{elmahdy2020patient} or by using synthetic deformations that mimic the one presented in the EMC dataset. 

In the paper, we compared our algorithm against different algorithms from various categories: non-learning (\texttt{elastix} \cite{elastix}, a popular conventional tool); hybrid \cite{MedPhys}, and GAN-based \cite{JrsGan}. The presented multi-task networks outperformed these approaches on the validation set and performed on par to these methods for the test set. However, the test time for the hybrid and \texttt{elastix} methods are in the order of minutes, while the presented methods have the advantage of fast prediction in less than a second. This enables online automatic re-contouring of daily scans for adaptive radiotherapy. Moreover, in our hybrid study \cite{MedPhys} we carried out an extensive dosimetric evaluation alongside the geometric evaluation. The predicted contours from that study met the dose coverage constraints in 86\%, 91\%, and 99\% of the cases for the prostate, seminal vesicles, and lymph nodes, respectively. Since our multi-task networks outperformed the geometrical results in that study, we expect that our contours would achieve a higher success rate in terms of the dose coverage. This could potentially reduce treatment related complications and therefore improve patient quality-of-life after treatment.

A promising direction for future research is the addition of a third task, potentially radiotherapy dose plan estimation. Hence, we can generate contours that are consistent with an optimal dose planning. Further studies could also focus on sophisticated MTL network architectures similar to sluice networks \cite{ruder2017sluice} or routing networks \cite{rosenbaum2017routing}. Moreover, we can study how to fuse the contours from the segmentation and registration paths in a smarter way rather than simply selecting one of them based on the validation set.


\section{Conclusion} \label{conclusion}
In this paper, we propose to formulate the registration and segmentation tasks as a multi-task learning problem. We presented various approaches in order to do so, both on an architectural level and via the loss function. We experimented with different network architectures in order to investigate the best setting that maximizes the information flow between these tasks. Moreover, we compared different loss weighting methods in order to optimally combine the losses from these tasks.

We proved that multi-task learning approaches outperform their single-task counterparts. Using an adaptive parameter sharing mechanism via Cross-stitch units gives the networks freedom to share information between these two tasks, which resulted in the best performance. An equal loss weighting approach had similar performance to more sophisticated methods. 

The cross stitch network with equal loss weights achieved a median MSD of 0.99 mm, 0.82 mm, 1.13 mm and 1.47 mm on the validation set and 1.09 mm, 1.24 mm, 1.02 mm, and 2.10 mm on the independent test set for the prostate, bladder, seminal vesicles, and rectum, respectively. That is equal or less than slice thickness (2 mm). Due to the fast inference of the methods, the proposed method is highly promising for automatic re-contouring of follow-up scans for adaptive radiotherapy, potentially reducing treatment related complications and therefore improving patient quality-of-life after treatment.

\section{Acknowledgment}
The HMC dataset with contours was collected at Haukeland University Hospital, Bergen, Norway, and was provided to us by responsible oncologist Svein Inge Helle and physicist Liv Bolstad Hysing.
The EMC dataset with contours was collected at Erasmus University Medical Center, Rotterdam, The Netherlands, and was provided to us by radiation therapist Luca Incrocci and physicist Mischa Hoogeman. They are gratefully acknowledged.
\bibliographystyle{ieeetr}
\bibliography{references}

\begin{thebibliography}{10}

\bibitem{nilashi2020disease}
M.~Nilashi, N.~Ahmadi, S.~Samad, L.~Shahmoradi, H.~Ahmadi, O.~Ibrahim,
  S.~Asadi, R.~Abdullah, R.~A. Abumalloh, and E.~Yadegaridehkordi, ``Disease
  diagnosis using machine learning techniques: A review and classification,''
  {\em Journal of Soft Computing and Decision Support Systems}, vol.~7, no.~1,
  pp.~19--30.

\bibitem{shen2017deep}
D.~Shen, G.~Wu, and H.-I. Suk, ``Deep learning in medical image analysis,''
  {\em Annual review of biomedical engineering}, vol.~19, pp.~221--248, 2017.

\bibitem{rueckert2014registration}
D.~Rueckert and J.~A. Schnabel, ``Registration and segmentation in medical
  imaging,'' in {\em Registration and Recognition in Images and Videos},
  pp.~137--156, Springer, 2014.

\bibitem{huo20193d}
Y.~Huo, Z.~Xu, Y.~Xiong, K.~Aboud, P.~Parvathaneni, S.~Bao, C.~Bermudez, S.~M.
  Resnick, L.~E. Cutting, and B.~A. Landman, ``3d whole brain segmentation
  using spatially localized atlas network tiles,'' {\em NeuroImage}, vol.~194,
  pp.~105--119, 2019.

\bibitem{wang2014multi}
H.~Wang, Y.~Cao, and T.~Syeda-Mahmood, ``Multi-atlas segmentation with
  learning-based label fusion,'' in {\em International Workshop on Machine
  Learning in Medical Imaging}, pp.~256--263, Springer, 2014.

\bibitem{MedPhys}
M.~S. Elmahdy, T.~Jagt, R.~T. Zinkstok, Y.~Qiao, R.~Shahzad, H.~Sokooti,
  S.~Yousefi, L.~Incrocci, C.~Marijnen, M.~Hoogeman, {\em et~al.}, ``Robust
  contour propagation using deep learning and image registration for online
  adaptive proton therapy of prostate cancer,'' {\em Medical physics}, vol.~46,
  no.~8, pp.~3329--3343, 2019.

\bibitem{JrsGan}
M.~S. Elmahdy, J.~M. Wolterink, H.~Sokooti, I.~I{\v{s}}gum, and M.~Staring,
  ``Adversarial optimization for joint registration and segmentation in
  prostate ct radiotherapy,'' in {\em International Conference on Medical Image
  Computing and Computer-Assisted Intervention}, pp.~366--374, Springer, 2019.

\bibitem{mahapatra2015joint}
D.~Mahapatra, Z.~Li, F.~Vos, and J.~Buhmann, ``Joint segmentation and groupwise
  registration of cardiac dce mri using sparse data representations,'' in {\em
  2015 IEEE 12th International Symposium on Biomedical Imaging (ISBI)},
  pp.~1312--1315, IEEE, 2015.

\bibitem{woerner2017evaluation}
A.~J. Woerner, M.~Choi, M.~M. Harkenrider, J.~C. Roeske, and M.~Surucu,
  ``Evaluation of deformable image registration-based contour propagation from
  planning ct to cone-beam ct,'' {\em Technology in cancer research \&
  treatment}, vol.~16, no.~6, pp.~801--810, 2017.

\bibitem{gu2013contour}
X.~Gu, B.~Dong, J.~Wang, J.~Yordy, L.~Mell, X.~Jia, and S.~B. Jiang, ``A
  contour-guided deformable image registration algorithm for adaptive
  radiotherapy,'' {\em Physics in Medicine \& Biology}, vol.~58, no.~6,
  p.~1889, 2013.

\bibitem{hansen2006repeat}
E.~K. Hansen, M.~K. Bucci, J.~M. Quivey, V.~Weinberg, and P.~Xia, ``Repeat ct
  imaging and replanning during the course of imrt for head-and-neck cancer,''
  {\em International Journal of Radiation Oncology* Biology* Physics}, vol.~64,
  no.~2, pp.~355--362, 2006.

\bibitem{brock2019adaptive}
K.~K. Brock, ``Adaptive radiotherapy: moving into the future,'' in {\em
  Seminars in radiation oncology}, vol.~29, p.~181, NIH Public Access, 2019.

\bibitem{sonke2019adaptive}
J.-J. Sonke, M.~Aznar, and C.~Rasch, ``Adaptive radiotherapy for anatomical
  changes,'' in {\em Seminars in radiation oncology}, vol.~29, pp.~245--257,
  Elsevier, 2019.

\bibitem{beljaards2020cross}
L.~Beljaards, M.~S. Elmahdy, F.~Verbeek, and M.~Staring, ``A cross-stitch
  architecture for joint registration and segmentation in adaptive
  radiotherapy,'' {\em arXiv preprint arXiv:2004.08122}, 2020.

\bibitem{lu2011integrated}
C.~Lu, S.~Chelikani, X.~Papademetris, J.~P. Knisely, M.~F. Milosevic, Z.~Chen,
  D.~A. Jaffray, L.~H. Staib, and J.~S. Duncan, ``An integrated approach to
  segmentation and nonrigid registration for application in image-guided pelvic
  radiotherapy,'' {\em Medical Image Analysis}, vol.~15, no.~5, pp.~772--785,
  2011.

\bibitem{pohl2006bayesian}
K.~M. Pohl, J.~Fisher, W.~E.~L. Grimson, R.~Kikinis, and W.~M. Wells, ``A
  bayesian model for joint segmentation and registration,'' {\em NeuroImage},
  vol.~31, no.~1, pp.~228--239, 2006.

\bibitem{yezzi2003variational}
A.~Yezzi, L.~Z{\"o}llei, and T.~Kapur, ``A variational framework for
  integrating segmentation and registration through active contours,'' {\em
  Medical image analysis}, vol.~7, no.~2, pp.~171--185, 2003.

\bibitem{unal2005coupled}
G.~Unal and G.~Slabaugh, ``Coupled pdes for non-rigid registration and
  segmentation,'' in {\em 2005 IEEE Computer Society Conference on Computer
  Vision and Pattern Recognition (CVPR'05)}, vol.~1, pp.~168--175, IEEE, 2005.

\bibitem{fu2020deep}
Y.~Fu, Y.~Lei, T.~Wang, W.~J. Curran, T.~Liu, and X.~Yang, ``Deep learning in
  medical image registration: a review,'' {\em Physics in Medicine \& Biology},
  2020.

\bibitem{yousefi2020esophageal}
S.~Yousefi, H.~Sokooti, M.~S. Elmahdy, I.~M. Lips, M.~T.~M. Shalmani, R.~T.
  Zinkstok, F.~J. Dankers, and M.~Staring, ``Esophageal tumor segmentation in
  ct images using a 3d convolutional neural network,'' {\em arXiv preprint
  arXiv:2012.03242}, 2020.

\bibitem{liu2019automatic}
C.~Liu, S.~J. Gardner, N.~Wen, M.~A. Elshaikh, F.~Siddiqui, B.~Movsas, and
  I.~J. Chetty, ``Automatic segmentation of the prostate on ct images using
  deep neural networks (dnn),'' {\em International Journal of Radiation
  Oncology* Biology* Physics}, vol.~104, no.~4, pp.~924--932, 2019.

\bibitem{kiljunen2020deep}
T.~Kiljunen, S.~Akram, J.~Niemel{\"a}, E.~L{\"o}yttyniemi, J.~Sepp{\"a}l{\"a},
  J.~Heikkil{\"a}, K.~Vuolukka, O.-S. K{\"a}{\"a}ri{\"a}inen, V.-P.
  Heikkil{\"a}, K.~Lehti{\"o}, {\em et~al.}, ``A deep learning-based automated
  ct segmentation of prostate cancer anatomy for radiation therapy planning-a
  retrospective multicenter study,'' {\em Diagnostics}, vol.~10, no.~11,
  p.~959, 2020.

\bibitem{cao2018deep}
X.~Cao, J.~Yang, L.~Wang, Z.~Xue, Q.~Wang, and D.~Shen, ``Deep learning based
  inter-modality image registration supervised by intra-modality similarity,''
  in {\em International Workshop on Machine Learning in Medical Imaging},
  pp.~55--63, Springer, 2018.

\bibitem{leger2020cross}
J.~L{\'e}ger, E.~Brion, P.~Desbordes, C.~De~Vleeschouwer, J.~A. Lee, and
  B.~Macq, ``Cross-domain data augmentation for deep-learning-based male pelvic
  organ segmentation in cone beam {CT},'' {\em Applied Sciences}, vol.~10,
  no.~3, p.~1154, 2020.

\bibitem{elmahdy2020patient}
M.~S. Elmahdy, T.~Ahuja, U.~A. van~der Heide, and M.~Staring,
  ``Patient-specific finetuning of deep learning models for adaptive
  radiotherapy in prostate ct,'' in {\em 2020 IEEE 17th International Symposium
  on Biomedical Imaging (ISBI)}, pp.~577--580, IEEE, 2020.

\bibitem{sokooti20193d}
H.~Sokooti, B.~de~Vos, F.~Berendsen, M.~Ghafoorian, S.~Yousefi, B.~P.
  Lelieveldt, I.~Isgum, and M.~Staring, ``3d convolutional neural networks
  image registration based on efficient supervised learning from artificial
  deformations,'' {\em arXiv preprint arXiv:1908.10235}, 2019.

\bibitem{tschandl2019comparison}
P.~Tschandl, N.~Codella, B.~N. Akay, G.~Argenziano, R.~P. Braun, H.~Cabo,
  D.~Gutman, A.~Halpern, B.~Helba, R.~Hofmann-Wellenhof, {\em et~al.},
  ``Comparison of the accuracy of human readers versus machine-learning
  algorithms for pigmented skin lesion classification: an open, web-based,
  international, diagnostic study,'' {\em The lancet oncology}, vol.~20, no.~7,
  pp.~938--947, 2019.

\bibitem{ardila2019end}
D.~Ardila, A.~P. Kiraly, S.~Bharadwaj, B.~Choi, J.~J. Reicher, L.~Peng, D.~Tse,
  M.~Etemadi, W.~Ye, G.~Corrado, {\em et~al.}, ``End-to-end lung cancer
  screening with three-dimensional deep learning on low-dose chest computed
  tomography,'' {\em Nature medicine}, vol.~25, no.~6, pp.~954--961, 2019.

\bibitem{hu2019observational}
L.~Hu, D.~Bell, S.~Antani, Z.~Xue, K.~Yu, M.~P. Horning, N.~Gachuhi, B.~Wilson,
  M.~S. Jaiswal, B.~Befano, {\em et~al.}, ``An observational study of deep
  learning and automated evaluation of cervical images for cancer screening,''
  {\em JNCI: Journal of the National Cancer Institute}, vol.~111, no.~9,
  pp.~923--932, 2019.

\bibitem{maidens2018artificial}
J.~Maidens and N.~B. Slamon, ``Artificial intelligence detects pediatric heart
  murmurs with cardiologist-level accuracy,'' {\em Circulation}, vol.~138,
  no.~Suppl\_1, pp.~A12591--A12591, 2018.

\bibitem{mak2019use}
R.~H. Mak, M.~G. Endres, J.~H. Paik, R.~A. Sergeev, H.~Aerts, C.~L. Williams,
  K.~R. Lakhani, and E.~C. Guinan, ``Use of crowd innovation to develop an
  artificial intelligence--based solution for radiation therapy targeting,''
  {\em JAMA oncology}, vol.~5, no.~5, pp.~654--661, 2019.

\bibitem{hu2018label}
Y.~Hu, M.~Modat, E.~Gibson, N.~Ghavami, E.~Bonmati, C.~M. Moore, M.~Emberton,
  J.~A. Noble, D.~C. Barratt, and T.~Vercauteren, ``Label-driven
  weakly-supervised learning for multimodal deformable image registration,'' in
  {\em 2018 IEEE 15th International Symposium on Biomedical Imaging (ISBI
  2018)}, pp.~1070--1074, IEEE, 2018.

\bibitem{mahapatra2018joint}
D.~Mahapatra, Z.~Ge, S.~Sedai, and R.~Chakravorty, ``Joint registration and
  segmentation of xray images using generative adversarial networks,'' in {\em
  International Workshop on Machine Learning in Medical Imaging}, pp.~73--80,
  Springer, 2018.

\bibitem{xu2019deepatlas}
Z.~Xu and M.~Niethammer, ``Deepatlas: Joint semi-supervised learning of image
  registration and segmentation,'' in {\em International Conference on Medical
  Image Computing and Computer-Assisted Intervention}, pp.~420--429, Springer,
  2019.

\bibitem{estienne2019u}
T.~Estienne, M.~Vakalopoulou, S.~Christodoulidis, E.~Battistela, M.~Lerousseau,
  A.~Carre, G.~Klausner, R.~Sun, C.~Robert, S.~Mougiakakou, {\em et~al.},
  ``U-resnet: Ultimate coupling of registration and segmentation with deep
  nets,'' in {\em International Conference on Medical Image Computing and
  Computer-Assisted Intervention}, pp.~310--319, Springer, 2019.

\bibitem{liu2020jssr}
F.~Liu, J.~Cai, Y.~Huo, L.~Lu, and A.~P. Harrison, ``Jssr: A joint synthesis,
  segmentation, and registration system for 3d multi-modal image alignment of
  large-scale pathological ct scans,'' {\em arXiv preprint arXiv:2005.12209},
  2020.

\bibitem{ronneberger2015u}
O.~Ronneberger, P.~Fischer, and T.~Brox, ``U-net: Convolutional networks for
  biomedical image segmentation,'' in {\em International Conference on Medical
  image computing and computer-assisted intervention}, pp.~234--241, Springer,
  2015.

\bibitem{fan2019birnet}
J.~Fan, X.~Cao, P.-T. Yap, and D.~Shen, ``Birnet: Brain image registration
  using dual-supervised fully convolutional networks,'' {\em Medical image
  analysis}, vol.~54, pp.~193--206, 2019.

\bibitem{nair2010rectified}
V.~Nair and G.~E. Hinton, ``Rectified linear units improve restricted boltzmann
  machines,'' in {\em ICML}, 2010.

\bibitem{pmlr-v37-ioffe15}
S.~Ioffe and C.~Szegedy, ``Batch normalization: Accelerating deep network
  training by reducing internal covariate shift,'' in {\em Proceedings of the
  32nd International Conference on Machine Learning} (F.~Bach and D.~Blei,
  eds.), vol.~37 of {\em Proceedings of Machine Learning Research}, (Lille,
  France), pp.~448--456, PMLR, 07--09 Jul 2015.

\bibitem{baxter2000model}
J.~Baxter, ``A model of inductive bias learning,'' {\em Journal of artificial
  intelligence research}, vol.~12, pp.~149--198, 2000.

\bibitem{meyerson2018pseudo}
E.~Meyerson and R.~Miikkulainen, ``Pseudo-task augmentation: From deep
  multitask learning to intratask sharing—and back,'' in {\em International
  Conference on Machine Learning}, pp.~3511--3520, PMLR, 2018.

\bibitem{abu1990learning}
Y.~S. Abu-Mostafa, ``Learning from hints in neural networks,'' {\em Journal of
  complexity}, vol.~6, no.~2, pp.~192--198, 1990.

\bibitem{zhang2017survey}
Y.~Zhang and Q.~Yang, ``A survey on multi-task learning,'' {\em arXiv preprint
  arXiv:1707.08114}, 2017.

\bibitem{misra2016cross}
I.~Misra, A.~Shrivastava, A.~Gupta, and M.~Hebert, ``Cross-stitch networks for
  multi-task learning,'' in {\em Proceedings of the IEEE Conference on Computer
  Vision and Pattern Recognition}, pp.~3994--4003, 2016.

\bibitem{chen2018gradnorm}
Z.~Chen, V.~Badrinarayanan, C.-Y. Lee, and A.~Rabinovich, ``Gradnorm: Gradient
  normalization for adaptive loss balancing in deep multitask networks,'' in
  {\em International Conference on Machine Learning}, pp.~794--803, 2018.

\bibitem{kendall2018multi}
A.~Kendall, Y.~Gal, and R.~Cipolla, ``Multi-task learning using uncertainty to
  weigh losses for scene geometry and semantics,'' in {\em Proceedings of the
  IEEE conference on computer vision and pattern recognition}, pp.~7482--7491,
  2018.

\bibitem{liu2019end}
S.~Liu, E.~Johns, and A.~J. Davison, ``End-to-end multi-task learning with
  attention,'' in {\em Proceedings of the IEEE Conference on Computer Vision
  and Pattern Recognition}, pp.~1871--1880, 2019.

\bibitem{elastix}
S.~Klein, M.~Staring, K.~Murphy, M.~A. Viergever, and J.~P. Pluim, ``Elastix: a
  toolbox for intensity-based medical image registration,'' {\em IEEE
  transactions on medical imaging}, vol.~29, no.~1, pp.~196--205, 2009.

\bibitem{abadi2016tensorflow}
M.~Abadi, A.~Agarwal, P.~Barham, E.~Brevdo, Z.~Chen, C.~Citro, G.~S. Corrado,
  A.~Davis, J.~Dean, M.~Devin, {\em et~al.}, ``Tensorflow: Large-scale machine
  learning on heterogeneous distributed systems,'' {\em arXiv preprint
  arXiv:1603.04467}, 2016.

\bibitem{liu2019variance}
L.~Liu, H.~Jiang, P.~He, W.~Chen, X.~Liu, J.~Gao, and J.~Han, ``On the variance
  of the adaptive learning rate and beyond,'' {\em arXiv preprint
  arXiv:1908.03265}, 2019.

\bibitem{qiao2017fast}
Y.~Qiao, {\em Fast optimization methods for image registration in adaptive
  radiation therapy}.
\newblock PhD thesis, Ph. D. thesis, Leiden University Medical Center, 2017.

\bibitem{ruder2017sluice}
S.~Ruder, J.~Bingel, I.~Augenstein, and A.~S{\o}gaard, ``Sluice networks:
  Learning what to share between loosely related tasks,'' {\em arXiv preprint
  arXiv:1705.08142}, vol.~2, 2017.

\bibitem{rosenbaum2017routing}
C.~Rosenbaum, T.~Klinger, and M.~Riemer, ``Routing networks: Adaptive selection
  of non-linear functions for multi-task learning,'' {\em arXiv preprint
  arXiv:1711.01239}, 2017.

\end{thebibliography}

\end{document}